\documentclass[acmsmall]{acmart}
\usepackage{multirow}

\usepackage{algorithm}
\usepackage{algpseudocode}
\usepackage{xspace}
\usepackage[utf8]{inputenc}
\usepackage{xcolor}
\usepackage{balance}  
\usepackage{mathtools}
\usepackage{url}
\usepackage{amsfonts}
\usepackage{textcomp}
\usepackage{algorithm}
\usepackage{multirow}
\usepackage{color}
\usepackage{soul}
\usepackage{subcaption}
\usepackage{enumitem}
\usepackage{array}

\usepackage{graphicx}

\newcolumntype{L}[1]{>{\raggedright\let\newline\\\arraybackslash\hspace{0pt}}m{#1}}
\newcolumntype{C}[1]{>{\centering\let\newline\\\arraybackslash\hspace{0pt}}m{#1}}
\newcolumntype{R}[1]{>{\raggedleft\let\newline\\\arraybackslash\hspace{0pt}}m{#1}}

\usepackage[normalem]{ulem}
\usepackage{color}
\usepackage{xcolor}

\usepackage{CJKutf8}

\newcommand{\system}{\emph{HearHere}\xspace}
\AtBeginDocument{%
  \providecommand\BibTeX{{%
    \normalfont B\kern-0.5em{\scshape i\kern-0.25em b}\kern-0.8em\TeX}}}

 \providecommand\BibTeX{{%
  Bib\TeX}}

\setcopyright{rightsretained}
\acmJournal{PACMHCI}
\acmYear{2024} \acmVolume{8} \acmNumber{CSCW1} \acmArticle{63} \acmMonth{4}\acmDOI{10.1145/3637340}

\makeatletter
\gdef\@copyrightpermission{
  \begin{minipage}{0.2\columnwidth}
   \href{https://creativecommons.org/licenses/by/4.0/}{\includegraphics[width=0.90\textwidth]{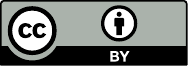}}
  \end{minipage}\hfill
  \begin{minipage}{0.8\columnwidth}
   \href{https://creativecommons.org/licenses/by/4.0/}{This work is licensed under a Creative Commons Attribution International 4.0 License.}
  \end{minipage}
  \vspace{5pt}
}
\makeatother

\begin{document}

\title{HearHere: Mitigating Echo Chambers in News Consumption through an AI-based Web System}

\author{Youngseung Jeon}
\email{ysj@ucla.edu}
\affiliation{%
 \institution{University of California Los Angeles}
 \country{USA}
}

\author{Jaehoon Kim}
\email{jaehoonkimm@hanyang.ac.kr}
\affiliation{%
  \institution{Hanyang University}
  \city{Seoul}
  \country{Republic of Korea}
}

\author{Sohyun Park}
\email{sally5004@ajou.ac.kr}
\affiliation{%
  \institution{Ajou University}
  \city{Suwon}
  \country{Republic of Korea}
}

\author{Yunyong Ko}
\email{yyko@cau.ac.kr}
\affiliation{%
  \institution{Chung-Ang University}
  \city{Seoul} 
  \country{Republic of Korea}
}

\author{Seongeun Ryu}
\email{ryuseong@hanyang.ac.kr}
\affiliation{%
  \institution{Hanyang University}
  \city{Seoul}
  \country{Republic of Korea}
}

\author{Sang-Wook Kim}
\email{wook@hanyang.ac.kr}
\affiliation{%
  \institution{Hanyang University}
  \city{Seoul}
  \country{Republic of Korea}
}

\author{Kyungsik Han}
\authornote{Corresponding author}
\email{kyungsikhan@hanyang.ac.kr}
\affiliation{%
  \institution{Hanyang University}
  \city{Seoul}
  \country{Republic of Korea}
}

\renewcommand{\shortauthors}{Youngseung Jeon et al.}
\renewcommand{\shorttitle}{HearHere}

\begin{abstract}
Considerable efforts are currently underway to mitigate the negative impacts of echo chambers, such as increased susceptibility to fake news and resistance towards accepting scientific evidence. Prior research has presented the development of computer systems that support the consumption of news information from diverse political perspectives to mitigate the echo chamber effect. However, existing studies still lack the ability to effectively support the key processes of news information consumption and quantitatively identify a political stance towards the information. 
In this paper, we present \system, an AI-based web system designed to help users accommodate information and opinions from diverse perspectives. \system facilitates the key processes of news information consumption through two visualizations. 
Visualization 1 provides political news with quantitative political stance information, derived from our graph-based political classification model, and users can experience diverse perspectives (\textit{Hear}). Visualization 2 allows users to express their opinions on specific political issues in a comment form and observe the position of their own opinions relative to pro-liberal and pro-conservative comments presented on a map interface (\textit{Here}). Through a user study with 94 participants, we demonstrate the feasibility of \system in supporting the consumption of information from various perspectives. Our findings highlight the importance of providing political stance information and quantifying users' political status to mitigate political polarization. In addition, we propose design implications for system development, including the consideration of demographics such as political interest and providing users with initiatives.
\end{abstract}



\received{January 2023}
\received[revised]{July 2023}
\received[accepted]{November 2023}
\maketitle

\section{Introduction}\label{Intro}

In a democratic society, the accurate consumption of information and the cultivation of the ability to do so are important. This practice can lead to more rational decision-making that is not heavily influenced by specific opinions or positions~\cite{fleming2014media, dahlberg2001internet, fishkin2005experimenting}. As the Internet is a primary source of information for many people and the volume of online information is immense, effectively helping people consume and share information from diverse perspectives is necessary but challenging~\cite{zhu2021context, lucero2017safe}. Researchers have proposed various support methods for this, including the development and use of computer technology. In particular, artificial intelligence (AI)-based recommendation systems have been designed to support efficient information consumption by learning users' demographic characteristics or online activity patterns and providing tailored information based on their preferences~\cite{shu2017fake}.

Although computer technology plays an important role in enabling people to access and share online information, it should be noted that providing information solely based on individuals' preferences and tendencies can inadvertently contribute to the formation of \textit{echo chambers}~\cite{shu2017fake}, a phenomenon where individuals are exposed primarily to the like-minded groups or information, leading to a reinforcement of shared narratives~\cite{garrett2011resisting}. Research has shown that echo chambers can have many negative outcomes, including the creation and dissemination of biased information~\cite{shu2017fake}, increased susceptibility to fake news~\cite{calvillo2020political, garrett2021conservatives}, resistance towards accepting scientific evidence~\cite{nisbet2015partisan}, and the adoption of unbalanced perspectives~\cite{hayes2018ideological}.

To prevent users from becoming polarized towards a specific political stance, many studies have proposed the use of computer-based tools designed to present information from diverse perspectives~\cite{liao2014can, gillani2018me, kriplean2012supporting, nelimarkka2019re}. Examples include ConsiderIt~\cite{kriplean2012supporting}, Balancer~\cite{munson2013encouraging}, and StarryThoughts~\cite{kim2021starrythoughts}. Recently, many studies have proposed AI models to help users quantitatively understand their information consumption. For example, Garimella et al.~\cite{garimella2017reducing} presented a model based on a recently developed user-level controversy score. Li and Goldwasser~\cite{li2021mean} proposed MEAN (Multi-head Entity Aware Attention Network), which measures news polarization by analyzing the relationship between the entities in news and external information.

\begin{figure*}
\centering

  \includegraphics[width=1\columnwidth]{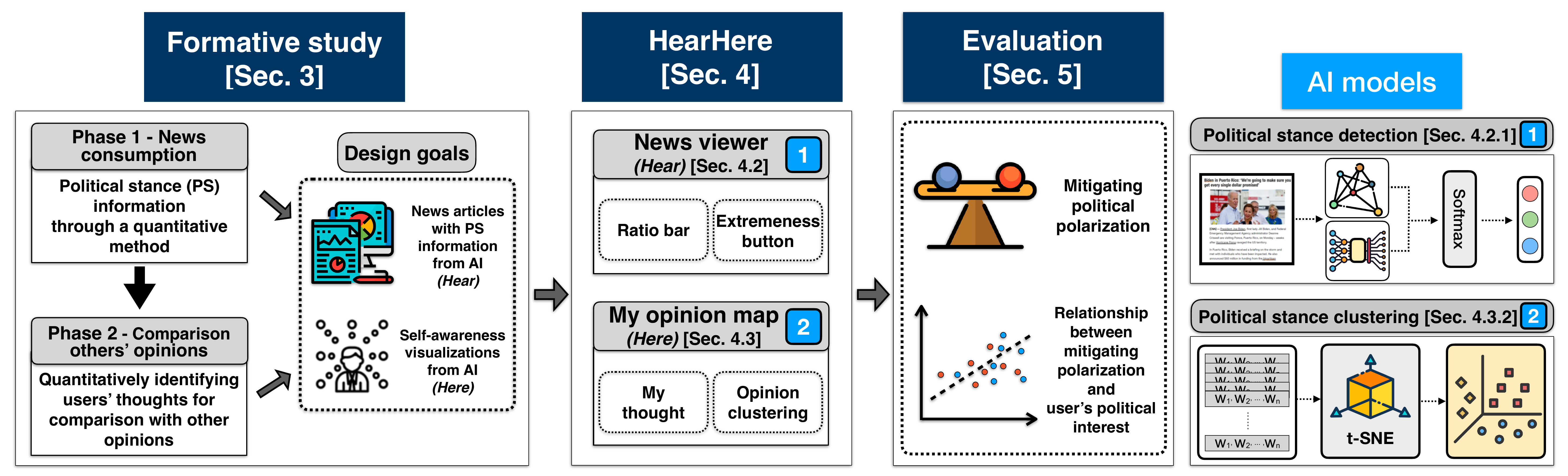}
  \vspace{-0.4cm}
  \caption{The overview of our research procedure.}
  \vspace{-0.5cm}
  ~\label{fig:Procedure}
\end{figure*}

However, prior research on the development of computer-based tools has been somewhat limited in a way of supporting news consumption. Digital literacy, which refers to one's ability to find, evaluate, and communicate information on digital platforms~\cite{gilster1997digital}, requires not only (1) understanding news information from official news sources (\textit{information understanding}~\cite{munson2013encouraging, liao2014can}) but also (2) comparing and confirming one's opinions relative to those of others from online communities (\textit{opinion comparison/confirmation}~\cite{kriplean2012supporting, nelimarkka2019re, kim2021starrythoughts})~\cite{fleming2014media}. 
However, existing studies have predominantly focused on supporting diverse perspectives in only one aspect of the information consumption process. Providing partial support in this process may not give people enough opportunities to digest information from diverse standpoints and could lead them to unbalanced news consumption. This, in turn, may inadvertently contribute to the reinforcement of echo chambers and impede efforts to mitigate polarization. 

In addition, despite previous research using AI technology for balanced information consumption, there are limited quantitative approaches to AI in the context of political stance information, as well as insufficient efforts to design an AI-based computer-supported tool for facilitating balanced information consumption regarding political stances. It is worth noting that studies examining information consumption have pointed out the importance of identifying users' political stances and content to foster balanced information consumption. Employing a quantitative methodology, such as AI, to provide political stances could help individuals assess the \textit{diversity} of news articles they encounter on a daily basis. Consequently, this could make an online environment where users can consume news based on political stances, enabling them to understand the rationality or biases inherent in the information. This would contribute to the mitigation of the echo chamber effect~\cite{dahlberg2001internet}. Furthermore, presenting users with news articles that encompass opposite political stances according to quantitative standards would allow researchers to investigate ways to support news consumption from a balanced standpoint, informed by individuals' experiences and behaviors~\cite{liao2014can}. Although considerable efforts have been made to predict the political stances of news articles by domain experts, the manual nature of this process, requiring a substantial amount of time and effort, may be susceptible to human biases~\cite{jeon2021chamberbreaker}.

In this paper, we present \system, an AI-based web system, which facilitates the consumption of information from diverse political perspectives to mitigate the echo chambers. \system incorporates an AI model specifically designed to classify the political stances of news articles and opinions. It supports two phases of information consumption: consuming news information and comparing one's own opinions with those of others. To facilitate this process, \system provides two types of visualizations that help users access information and opinions from diverse perspectives. It is important to note that the primary objective of \system is not to alter users' own political stance but to make them aware of the importance of consuming information from diverse perspectives and to encourage them to consider such importance for their balanced understanding and consumption of news.

With \system, we strive to answer the following research questions (RQs):
\begin{itemize}
    \item RQ1: How can we incorporate key components of information consumption in the context of digital literacy into the design of \system?
    \item RQ2: Does \system help the participants realize the importance of information diversity?
    \item RQ3: How do changes in perception differ according to user characteristics? 

\end{itemize}

To answer the research questions, we conducted research as follows (Figure~\ref{fig:Procedure}):
\begin{itemize}
    \item \textit{User understanding} (Sec.~\ref{Formative}): Through interviews with six participants who regularly read news articles online, we obtained valuable insights into their existing news consumption practices, identified challenges, and found potential strategies to foster a more balanced news consumption experience. We confirmed the necessity of developing AI models capable of predicting the political stances of new articles based on quantitative standards. We also identified two fundamental phases of information consumption---the \textit{information check} phase, which involves consuming news information and the \textit{information comparison} phase, which entails comparing one's opinions with those of others. These two phases were found to occur interchangeably during the process of news consumption.
    \item \textit{HereHear development} (Sec.~\ref{HH}): \system consists of two primary functions, each corresponds to the phase of news consumption. The first function allows users to consume news articles with political stance information provided by an AI model. Users can have the flexibility to adjust the proportion of news articles based on the recommended level of political stances by the state-of-the-art political stance prediction model developed by our research team (Sec.~\ref{HH-1}). The second function allows users to explore the relationship between their opinions on specific issues and those expressed by representative conservative and liberal communities through a two-dimensional space map (Sec.~\ref{HH-2}).  
    \item \textit{User study} (Sec.~\ref{US}): To measure the effectiveness of \system, we conducted a user study with 94 participants. The participants were instructed to use \system for three days. They were asked to view updated news articles and user responses (comments) to the given news topics and to leave their own comments on the given news topics. In particular, we measured the differences in their responses to the echo chamber (EC) breaking questions (Table~\ref{tab:ECQuestion}), which aims to gauge their perceptions regarding their access and consumption of diverse news information before (pre-survey) and after (post-survey) using \system. 
\end{itemize}

Through our user study, we investigated if participants using \system would recognize the importance of information diversity and if their acceptance of diverse perspectives would vary based on their characteristics. We observed a significant increase in EC breaking scores, with a positive relationship between \system usage and these scores. Differences in the EC breaking scores were found among user groups based on age, political stance, and political interest. Notably, individuals with high political interest and strong stances exhibited a decrease in the EC breaking scores, indicating resistance to changing their beliefs when exposed to contrasting information.

Interestingly, the group characterized by high political interest and a strong political stance showed a decrease in the EC breaking scores, whereas all other groups showed an increase. This suggests that certain individuals may exhibit resistance to changing their perspectives and beliefs, even when exposed to information that contradicts their own stances. There is certainly no one-size-fits-all solution, especially to this complex social phenomenon. Nevertheless, our study results offer a promising design direction, as \system demonstrated its potential to promote reflection on the importance of information diversity. 
With a well-designed AI model capable of classifying the political stance of news articles, educational or information access systems like \system have the potential to support balanced news consumption and promote digital literacy. We also observed some limitations, such as the backfire effect of information diversity, different intentions, and motivations of participants in news consumption.

Notably, our study results shed light on the role and use of AI in quantitatively providing political information based on the participants' political interests. Participants were generally open and receptive to AI-generated results when they could be used as a quantitative standard for making one's decisions on specific political issues. However, participants who hold relatively strong political interests and stances showed higher resistance to the AI-generated results. In addition, participants expressed a desire for greater initiative and control over their interactions with AI, as well as customized options to tailor the system to their needs and interpretability. The purpose of this initiative varied based on the level of political interest, with the high political interest group seeking information aligning with their existing beliefs and the low political interest group desiring a broader range of opinions efficiently. These results highlight the importance of considering users' political stances and expectations when designing computer-supported tools that aim to facilitate balanced information consumption.

The contributions of this study are as follows:
\begin{itemize}
\vspace{-0.1cm}
    \item We develop \system to promote balanced news information consumption by understanding the key processes involved in online information consumption in terms of digital literacy.  
    \item We demonstrate the design rationales of \system through user studies in the real environment and identify that the approach to information consumption varies depending on one's political stance and political interest.
    \item We discuss important design implications for the computer tool that aims to support news consumption from diverse perspectives.
\end{itemize}

\section{RELATED WORK}
In this section, we review previous work on designing and building interfaces for mitigating an echo chamber effect. After that, we present the applicability of AI models in solving this problem to support diverse opinions based on quantitative information from AI. 

\subsection{Mitigating echo chambers through computer system design}
HCI communities have actively studied the development and design of social information systems for mitigating the echo chamber effect. In terms of mitigating echo chambers, digital literacy requires two aspects: (1) understanding news information and (2) comparing one's opinions with those of others in online communities by accessing diverse perspectives~\cite{fleming2014media}. To mitigate polarization, prior studies have presented systems designed to encourage people to consume information from diverse perspectives in these two aspects.

Regarding the first aspect, Munson et al.~\cite{munson2013encouraging} developed a widget that represents users' political biases based on the news they had read, helping them recognize their biased consumption statuses and encouraging them to read news articles from opposing perspectives. Liao and Fu~\cite{liao2014can} designed an interface that promotes users' consumption of information from diverse perspectives through visualizations of political stances (conservative or liberal) and extremeness (moderate or extreme). 

There are also studies that pertain to the second aspect. ConsiderIt~\cite{kriplean2012supporting} is a system that helps users effectively view particular political issues from diverse perspectives by organizing the pros and cons of the issues based on other people's opinions. 
Nelimarkka et al.~\cite{nelimarkka2019re} designed an interface that assists users in understanding different perspectives by exposing them to opposing viewpoints on social media. StarryThoughts~\cite{kim2021starrythoughts} is a system that helps users consume diverse opinions by providing them with functions to sort people's opinions on social issues based on demographic information and to navigate these opinions effectively. Polis~\cite{small2021polis} presented a system for supporting decision-making on situations such as policy development that require majority consensus. Rieger et al.\cite{rieger2021item} studied the effect of obfuscation (i.e., hiding the result unless the user clicks on it) with warning labels and the effect of task on interaction with attitude-confirming search results. 

Beyond prior studies that partially supported the key processes of digital literacy, \system includes both the processes of information consumption and opinion observation and comparison. Covering those processes is likely to mitigate the possibility of users falling into echo chambers or being politically polarized. This also means a potential extension and application of \system for use by more people and groups looking for ways to achieve digital literacy.

\subsection{AI to classify a political stance for mitigating echo chambers}

In political news consumption, the particular perspectives in political information can prime readers to take similar stances and shape their world view~\cite{gentzkow2011ideological}. Researchers place importance on helping people exposed to diverse perspectives to mitigate the echo chamber effect~\cite{liao2014can, gillani2018me, donkers2021dual, tommasel2021want}. Researchers define \textit{diversity} of information based on the relationship between the political stances of information consumers and news articles. For example, if the political stances of the user and the news article are different, they defined that the user is exposed to diverse perspectives~\cite{iyengar2009red, mullainathan2005market}. In other words, providing information consumers with political stance information through a quantitative methodology, such as AI, could be a starting point to mitigating echo chambers.

Much research has been done for the purpose of developing AI models to classify political stances on news information. Feng et al.~\cite{feng2021knowledge} presented KGAP, a political stance detection method that incorporates external domain knowledge to reflect the political context. Li et al.~\cite{li2019encoding} suggested an AI model based on Graph Convolutional Networks for representing relational information, to capture the documents’ social context. Li and Goldwasser~\cite{li2021mean} proposed to MEAN (Multi-head Entity Aware Attention Network) which calculates the relationship between the entities in news and external information to determine polarization. Sun et al.~\cite{sun2018stance} presented a neural model to fully employ various linguistic information to construct the document representation with an attention mechanism to protect the political stances of news articles.

At an application level, there has been much effort to make usage of AI models in building an interface to mitigate an echo chamber effect (e.g., AI to classify people’s sentiment on news articles~\cite{gao2018burst}, a model to recommend accounts to users~\cite{tommasel2021want, gillani2018me}, and an AI model to identify a type of echo chambers~\cite{donkers2021dual}). However, to the best of our knowledge, there is no research that applies an AI model to predict the political stances of news articles in building an interface aiming to mitigate echo chambers.

In this work, we propose using AI to predict the political stance of news articles as a tool to help users consume news articles in balanced information consumption. We developed a novel knowledge-aware approach to accurate political stance prediction~\cite{ko2023khan}, employing (1) hierarchical attention networks to learn the relationships among words/sentences at different levels and (2) knowledge encoding to incorporate external knowledge (both common and political) for real-world entities into the process of political stance prediction with outstanding performance. On top of that, we aim to provide an interactive function (Visualization 1, Sec.~\ref{HH-1}) that allows users to consume news articles with political stance information from our model.

\subsection{Self-awareness for mitigating echo chambers}
HCI researchers exploring how to mitigate the effects of polarization have proposed computer tools to raise users' self-awareness by allowing them to compare their thoughts with diverse opinions, including opposing opinions~\cite{munson2013encouraging, gao2018burst, kriplean2012supporting, faridani2010opinion, kim2021starrythoughts}. 
Balancer~\cite{munson2013encouraging} is a widget designed to nudge its users to read balanced political viewpoints. This widget represented the aggregate political lean of users' weekly and all-time reading behaviors to encourage those whose reading leaned one way to read news in a balanced perspective. 
Gao, Do, and Fu~\cite{gao2018burst} designed an intelligent system that aims to improve awareness of diverse social opinions by providing visual hints and recommendations of opinions (e.g. news articles and comments) on different sides with different indicators. ConsiderIt~\cite{kriplean2012supporting} is a system that helps users effectively view particular political issues from diverse perspectives by organizing pros and cons of the issues based on users' thoughts (self-awareness). Opinion Space~\cite{faridani2010opinion} is a web-based interface, incorporating ideas from deliberative polling, dimensionality reduction, and collaborative filtering, that
allows users to navigate through a diversity of comments based on users' thoughts. StarryThoughts~\cite{kim2021starrythoughts} is a system that helps users consume diverse opinions by providing them with functions to sort people's diverse opinions on social issues based on users' demographic information and thoughts.

Considerable efforts are being made to provide users with an opportunity for self-awareness with the data reflecting users' behavior, such as the characteristics of news articles users had consumed, demographic information, and survey score results regarding political issues~\cite{munson2013encouraging, jeon2021chamberbreaker, gao2018burst}. However, for all their efforts, using these data could be limited in reflecting their opinions and thoughts themselves. This leaves substantial room for the best method to reflect users' thoughts.

In this work, we suggest using users' thoughts themselves in text form for reflecting and presenting them intuitively by using AI models. In Visualization 2 (Sec.~\ref{HH-2}), users can submit their opinions about a specific political issue in the comment form and confirm the position of their own opinions between pro-liberal and pro-conservative comments presented in a map interface.

\subsection{Side effect of supporting diverse perspectives}
HCI researchers emphasize the importance of developing interfaces that support having diverse perspectives as a way to mitigate echo chamber effects. The aim is to keep users from becoming polarized to a particular stance by providing diversity in information consumption. However, studies in diverse domains, such as political science, psychology, and cognitive science, have pointed out that focusing too much on information diversity could lead to unexpected consequences---reinforcing people's beliefs and exacerbating political polarization~\cite{nyhan2010corrections, taber2006motivated}---, which is called a backfire effect.

In the field of psychology, Ecker et al.~\cite{ecker2022psychological} provided insights into the formation of false beliefs by elucidating the cognitive, social, and emotional factors that contribute to the endorsement of misinformation. Their study highlighted the psychological barriers to knowledge revision once misinformation is debunked, suggesting that corrections perceived as attacks on one's worldview can paradoxically reinforce misinformation, leading to a 'backfire' effect. Jasmyne et al.~\cite{sanderson2022listening} investigated the potential occurrence of 'familiarity backfires' under cognitive load conditions. Despite the potential enhancement of familiarity through the repetition of misinformation, the cognitive load was found to inhibit the integration of corrections, thereby reducing their effectiveness. This may lead to the emergence of a 'backfire effect'.

In cognitive science, Swire-Thompson et al.~\cite{swire2020searching} questioned the empirical validity of 'backfire effects' and called for stronger measures, robust experimental designs, and more robust theoretical connections to advance the field. Similarly, Ecker et al.~\cite{ecker2014people} proposed that 'backfire effects' may reflect individuals' attempts to defend and maintain their attitudes and beliefs, rather than indicating attitudinal change. Contrary to prevailing understanding, research across disciplines continues to expand the examination of the backfire effect and challenge conventional views.

In political science, Bakshy et al.~\cite{bakshy2015exposure} conducted an extensive examination of the interaction between approximately 10.1 million U.S. Facebook users and socially shared news. They argued that the potential for exposure to divergent perspectives on social media ultimately lies with individual users. Nyhan and Reifler~\cite{nyhan2010corrections} have pointed out that the stronger consumers’ beliefs are, the more likely they are to strengthen their beliefs and worsen political polarization when they are exposed to opposing information. Taber and Lodge~\cite{taber2006motivated} mentioned that people who are exposed to diverse information about their beliefs tend to refuse the information by motivated reasoning to emphasize the difference and increase devotion to existing beliefs. Ecker et al.~\cite{ecker2019political} suggested result that even when misinformation provided to information consumers is later retracted, it can have a backfire effect when the information aligns with their beliefs. Nyhan et al.~\cite{nyhan2023like} presented that the majority of content viewed on the platform originates from sources with similar viewpoints, with political information and news making up a relatively small portion of these interactions. González-Bailón \cite{gonzalez2023asymmetric} evaluated the political news users could potentially access in their feeds, the information presented to them following algorithmic curation, and the content they interacted with.

Given that previous research in the field of the design for mitigating echo chambers has aimed to increase exposure to diverse perspectives and promote awareness of these effects, we acknowledge to some extent that the backfire effect is an unavoidable limitation highlighted in previous relevant studies. Nevertheless, the backfire effect discussed in previous research seems to be related to the degree of political beliefs. The stronger one's beliefs, the more resistance arises to opposing information, leading to further support for one’s own beliefs. Previous research has shown that motivated social coordination~\cite{kruglanski1996motivated} or motivated reasoning~\cite{kahan2013cognitive} is one of the main causes for people to reject information with opposing views, regardless of whether the content is false or true. This suggests that individuals tend to accept or reject information based on the congruence between the information and their existing worldview, and are more likely to consume information that is consistent with their beliefs~\cite{swire2017processing}.

In this study, we conducted an investigation into the relationship between users' political interests and the backfire effect, focusing specifically on individuals with strong political interests. To the best of our knowledge, this research represents an early effort to highlight the importance of considering users' political interests in the development of an AI-based interface designed to effectively mitigate the formation of echo chambers, taking into account the backfire effect.

\begin{table}[]
\caption{Demographic information of formative study participants and the average daily time they spent on reading political news. We used $P_f^X$ to denote participant number X in the formative study in the ID column.}
\label{tab:participant-f}
\begin{tabular}{|c|c|c|c|c|}
\hline
\multicolumn{1}{|l|}{ID} & \multicolumn{1}{l|}{Gender} & \multicolumn{1}{l|}{Age} & \multicolumn{1}{l|}{Political stance} & \begin{tabular}[c]{@{}c@{}}Minutes Reading News Online\\ (daily)\end{tabular} \\ \hline
$P_f^1$                      & Male                        & 28                       & Conservative                          & 30-45                                                                     \\ \hline
$P_f^2$                        & Female                      & 36                       & Conservative                          &  35                                                                       \\ \hline
$P_f^3$                        & Male                        & 43                       & Conservative                          & 30-45                                                                     \\ \hline
$P_f^4$                        & Male                        & 42                       & Liberal                               & 30-40                                                                     \\ \hline
$P_f^5$                        & Female                      & 28                       & Liberal                               & 30-45                                                                        \\ \hline
$P_f^6$                        & Male                        & 36                       & Liberal                               & 40-60                                                                     \\ \hline
\end{tabular}

\end{table}

\section{Formative study} \label{Formative}
To gain insight from news consumers in the real world for balanced information consumption (RQ1), we conducted interviews with six participants who read news via online websites. We detail the process of political news information consumption with respect to the following aspects: 1) the process of political news information consumption, 2) the challenges in the existing news interface for balanced information consumption, and 3) possible solutions for those challenges. 
Each interview took an average of 30 minutes.

\subsection{Interview procedure}
We recruited six participants from university mailing lists and through the snowball sampling method~\cite{goodman1961snowball}. {Since the Pew Research Center reported that 34\% of the public said that they accessed the online to read the news on the previous day~\footnote{https://www.pewresearch.org/politics/2010/09/12/americans-spending-more-time-following-the-news}, we recruited participants who spend at least 35 minutes a day with online news. All participants (two females and four males) consume news information on the web (e.g., news web platforms and SNSs) on a daily basis. Considering the direction of our study, we kept the same ratio of the participants according to their political stance (three pro-conservatives and three pro-liberals). The age of the participants ranged from 20 to 40 years. Detailed information about the participants can be found in Table~\ref{tab:participant-f}.

The interviews were conducted via an online Zoom meeting on September 1-5, 2022, and each interview took about 60 minutes. Two researchers (the first and the second authors) conducted the interviews. The interviews were audio-recorded and transcribed for later analysis. Each participant was compensated with a \$15 gift card for their time. During the interview, each session took approximately 60 minutes and included a semi-structured interview followed by a feedback session on the general idea of consuming political news online.

We asked three questions as below.
\begin{itemize}
    \item How do you consume news online? (the two phases of political news information consumption)
    \item What are the barriers and challenges in balanced news information consumption?
    \item What are the potential solutions and coping strategies to address these challenges? 
\end{itemize}

\begin{table*}[]
\centering
\caption{Results of formative studies.}
\fontsize{8}{10}\selectfont

\renewcommand{\arraystretch}{1.1}
\begin{tabular}{  >{\raggedright\arraybackslash}p{4cm} | >{\raggedright\arraybackslash}p{5cm} | >{\raggedright\arraybackslash}p{3.5cm}  }
\hline
\multicolumn{1}{c|}{\textbf{\centering Questions in the formative study}} & \multicolumn{1}{c|}{\textbf{\centering Responses}} & \multicolumn{1}{c}{\textbf{\centering Insight}} \\ \hline
Q1. How do you consume news online? (current practices of news consumption) & Two phases of news consumption:\newline 1) news information consumption phase \newline 2) opinion comparison phase & To support users’ political news consumption in the real world, the interface needs to cover the whole process (i.e. two phases). \\ \hline
Q2. What are the barriers and challenges in balanced news information consumption? & 1) In the news information consumption phase, participants mentioned that it is difficult to consume information from a different standpoint. \newline 
2) In the comparison of others’ opinions phase, they mentioned that it is difficult to consume opposing opinions for balanced information consumption. & For the issue in each phase, we need to develop AI to \newline 1) quantitatively define political stances of news and \newline 2) quantitatively define their thoughts based on others’ thoughts.\\ \cline{1-2}
Q3. What are the potential solutions and coping strategies to address these challenges? & 1) In the news information consumption phase, people should receive quantitative information about political stances, defined by a quantitative method.\newline 2) In the opinion comparison phase, people should quantitatively identify their thoughts for comparison with others' opinions. & \\ \hline
\end{tabular}

\label{tab:FormativeResult}
\vspace{-0.2cm}
\end{table*}

Based on personal experience of online news consumption, we empirically recognized the two phases of political news information consumption. These phases correspond to the elements of digital literacy, namely understanding information and comparing one's opinion with others~\cite{fleming2014media}. Given that the primary objective of our research was to develop an AI-based interface that supports balanced news consumption by online news consumers in the real world, we sought to confirm "the two phases" of online news consumption from people who met the study conditions through the formative study.

After the interviews were completed, we applied thematic analysis and iterative open coding~\cite{smith2015qualitative} to analyze the interview transcripts. Two researchers who specialized in qualitative analysis coded and analyzed the transcripts for emerging themes. The findings were discussed iteratively among the co-authors of this paper until a consensus was reached.

\subsection{Results}
Below, we summarize the process of political news online consumption and current challenges and possible solutions that emerged from the interviews. When reporting interview quotes, we used $P_f^X$ to denote participant number X in the formative study. Detailed information about the results can be found in Table~\ref{tab:FormativeResult}.

\subsubsection{News information consumption online}
From the results, we identified the two steps that people select for news information consumption.

\begin{itemize}
    \item News information consumption: When participants found specific political issues, they searched for news on major news portals or checked recommended news by the system. After reading the news, they organized their opinions. 
    \item Comparison of others' opinions: Participants mostly go to online political communities they are interested in and regularly visit. While reading over others' opinions, they conformed to the same views/opinions that support a particular political party and are highly prevalent in the community. As they are more exposed to such activities and spend more time on the topics discussed by community members, they justify their thinking.
\end{itemize}

We found that ways of reading news articles vary by participant. 
Two participants answered that they do not read news articles from other news media that do not match their political stance, three answered they sometimes read news articles from the opposing stance, and one answered that she reads almost the same ratio for the news articles from two different political stances. We also noticed that the participants organized their thinking on a specific issue. When a specific political issue occurred, all participants showed a tendency to interpret the issue in advance based on their political stance and to organize their thoughts while reading the news articles. Two participants said that sometimes they change their minds when they read a news article that is different from their own point of view, but the other four participants answered that they do not usually change their minds.

As another practice for news consumption, the participants compared their own organized opinions to others' opinions. First, they read the comments from the news portal. They also visit an online community where they spend some amount of time every day and check others' opinions on the news. 
They mentioned that they sometimes write a comment or send a direct message to community members, sharing their own thoughts, asking questions, or discussing the issue that they are interested in. Not all participants in our formative study are active in their online community, but they said that they regularly visit their online community and check the atmosphere regarding the news topics shared in the community.

In summary, we identified two phases of news consumption in the real world through interviews. Those phases correspond to the elements of digital literacy (i.e., understanding information and comparing one's opinions with others)~\cite{fleming2014media}. 

This means the necessity of covering the key processes in mitigating the possibility of falling users into echo chambers or political polarization.

\subsubsection{Challenges and possible solutions}
We identified the challenges that users experience during the process of news information consumption and the possible solutions to address those challenges based on the two phases.

First, in the news information consumption phase, participants mentioned that it is difficult to consume information from a different standpoint, because new articles do not usually come with information about their political stance. 

In the case of the major news outlets, people could guess the political stances of the news articles, because large-scale news outlets often represent a particular political side. However, for people who read or happen to read news articles from minor new outlets may not be able to easily identify the political stances of the news articles. In this case, the number of news articles that people can identify their political stances can be limited; thus, a challenge exists in consuming new articles from diverse perspectives, even if people want to do. $P^{4}_{f}$ mentioned, ``\textit{We can easily click the news media that has a typical political stance from major news media, but sometimes cannot identify the political stance of the articles from the minor news media}''. $P^{3}_{f}$ said the following, ``\textit{If I can figure out the political stances of news articles before reading news, it would be helpful for me to consume the news having the opposing view. But the current site I visit regularly does not provide such a function}''. 

Participants feel difficulties that come from the limitation of quantitatively defining a political stance. This reveals the potential usefulness of quantitative information from AI to classify a political stance as an indicator when people consume political news.

Second, in the comparison of others' opinions phase, they mentioned that it is difficult to consume opposing opinions for balanced information consumption because there is a limitation to easily and efficiently consuming the opposing information. If participants compare their opinions with opposing opinions, they should first go to communities that support opposing political sides. However, it is unrealistic to check a lot of posts related to a particular political issue one by one. This prevents them from overall understanding the opposing opinion. For example, $P^{5}_{f}$ said ``\textit{Because political issues are difficult to talk about offline, there is no better place than the online community to share my opinion on specific political issues. However, sometimes there are too many opinions on political online communities to support a particular political side. In retrospect, I forgave to check opposing opinions many times due to this challenge}''. In addition, the participants mentioned that they could not recognize how biased they are for a particular political issue, because there is no standard to help them introspect themselves. For example, $P^{4}_{f}$ mentioned, ``\textit{I sometimes agree with a particular political idea. If I hadn't met my friend who has a natural political position on the issue, I wouldn't have known how biased I was.}'' Participants feel difficulties that come from the limitation of quantitatively comparing information from both political sides and identifying their political position. This reveals the potential usefulness of quantitative information to visualize the opinions of both sides on particular political issues and users' thoughts based on quantitative aids including AI models.

In summary, the participants responded that ambiguous qualitative criteria and guidelines in defining a political stance are key challenges preventing balanced information consumption. In this research, we strived to address such challenges through a design and development of a tool for providing news articles with political stances defined by a quantitative methodology and identifying a relationship between their thoughts and other opinions from both political sides.

\begin{figure}
\centering
  \includegraphics[width=0.85\columnwidth]{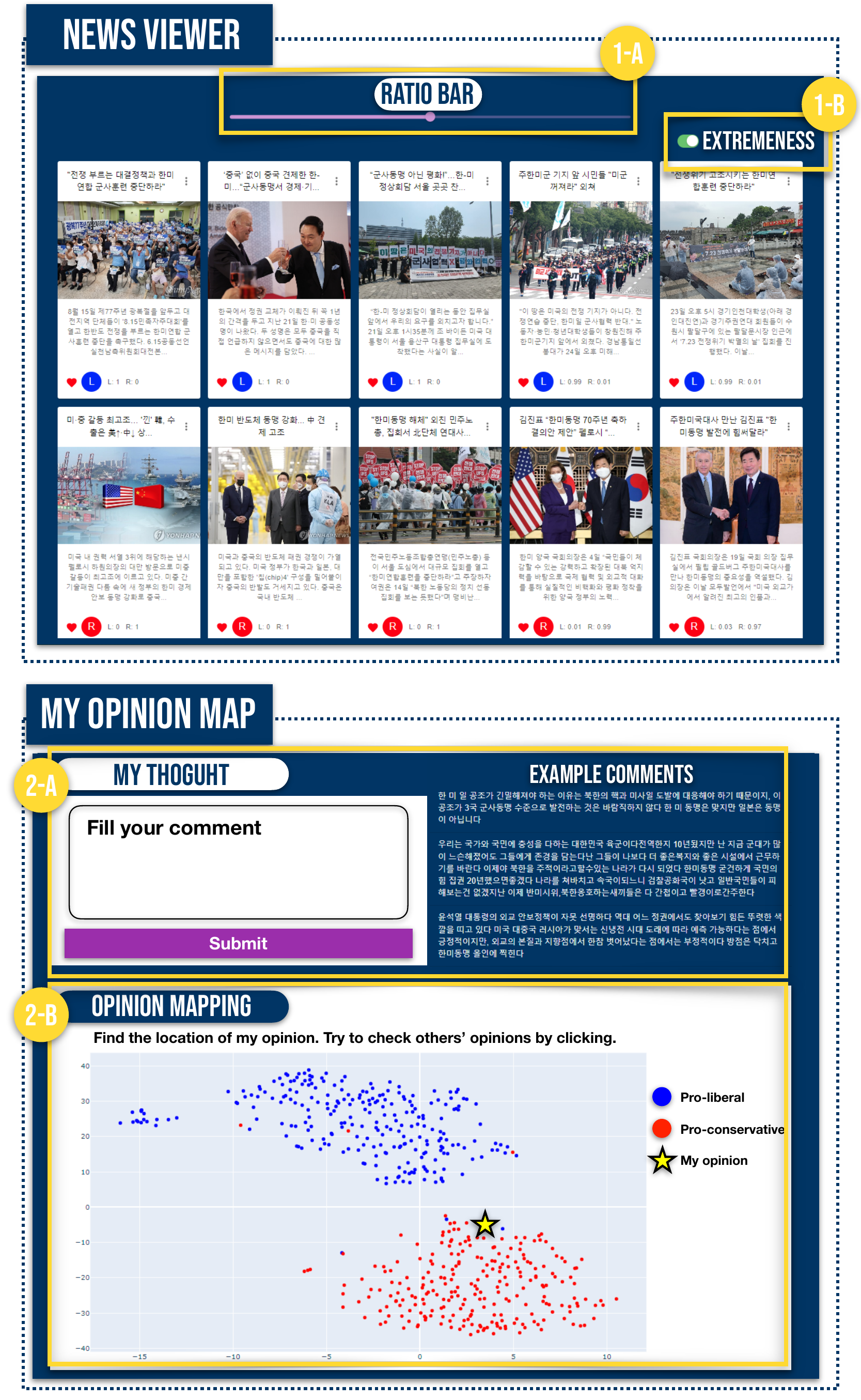}
  \vspace{-0.2cm}
  \caption{\system with two main interactive visualization interfaces: News viewer and My opinion map (some parts of the system were blurred to anonymize).}
\vspace{-0.6cm}
  ~\label{fig:HereHear}
\end{figure}

\begin{table*}[]
\centering
\caption{Design goals of the balanced news consumption considered in \system.}

\scriptsize
    \small
    \resizebox{\textwidth}{!}{
    \begin{tabular}{ p{2.5cm} p{3cm} p{6.6cm} p{2cm}}
    \toprule
    \multicolumn{1}{c}{\textbf{News consumption phase}} & \multicolumn{1}{c}{\textbf{Requirement}} & \multicolumn{1}{c}{\textbf{Goal}}  & \multicolumn{1}{c}{\textbf{Interface visualization}} \\ \midrule
    {\begin{tabular}[l]{@{}l@{}}Information\\  understanding\end{tabular}} & 
    {\begin{tabular}[l]{@{}l@{}}Political stance \\ information defined \\ by a quantitative \\ methodology\end{tabular}} &
    {\begin{tabular}[c]{@{}l@{}}Provides \textbf{news articles with their political } \\ \textbf{stance information through AI} \\(e.g., political stance and extremeness)\end{tabular}} 
    & \textit{News viewer - Hear (Sec.~\ref{HH-1})}
    \\\midrule
    Opinion comparison/confirmation &
    {\begin{tabular}[l]{@{}l@{}}Quantitatively \\ identifying users' \\ thoughts for \\comparison wtih\\ other opinions\end{tabular}}  &
    {\begin{tabular}[c]{@{}l@{}}Provides \textbf{self-awareness visualizations} \\ \textbf{through AI} for comparing \\ 
    between user's thoughts and both opinions\\ 
    supporting  particular political sides
    \\ (conservative or liberal)
    \end{tabular}}&

    \textit{My opinion map - Here (Sec.~\ref{HH-2})}
    \\
    \bottomrule
    \end{tabular} 
    }
    \label{tab:goal_step}
    \vspace{-0.2cm}
\end{table*}

\section{System development: HEARHERE}\label{HH}

\subsection{System design goals}\label{goals}
Based on design goals derived from a formative study, we had two major goals for the design of \system (Table~\ref{tab:goal_step}). Goal 1 provides news articles with their political stance information through AI (e.g., political stance and extremeness). Goal 2 provides self-awareness visualizations through AI for comparing users' thoughts and both opinions supporting  particular political sides. Users can locate their own thoughts among the opinions of the two political sides on the map and check the political relationship between their thoughts and those of others. 

Based on two goals, \system provides two main visualizations: \textit{News viewer} and \textit{My opinion map} (Figure~\ref{fig:HereHear}). With \system, people can consume news information with quantitative political-related information, such as political stance and level of political extremeness, and have the opportunity to express their viewpoints regarding a particular political matter and determine the alignment of their opinions among the pro-liberal and pro-conservative comments.

\begin{figure}
\centering
  \includegraphics[width=0.7\columnwidth]{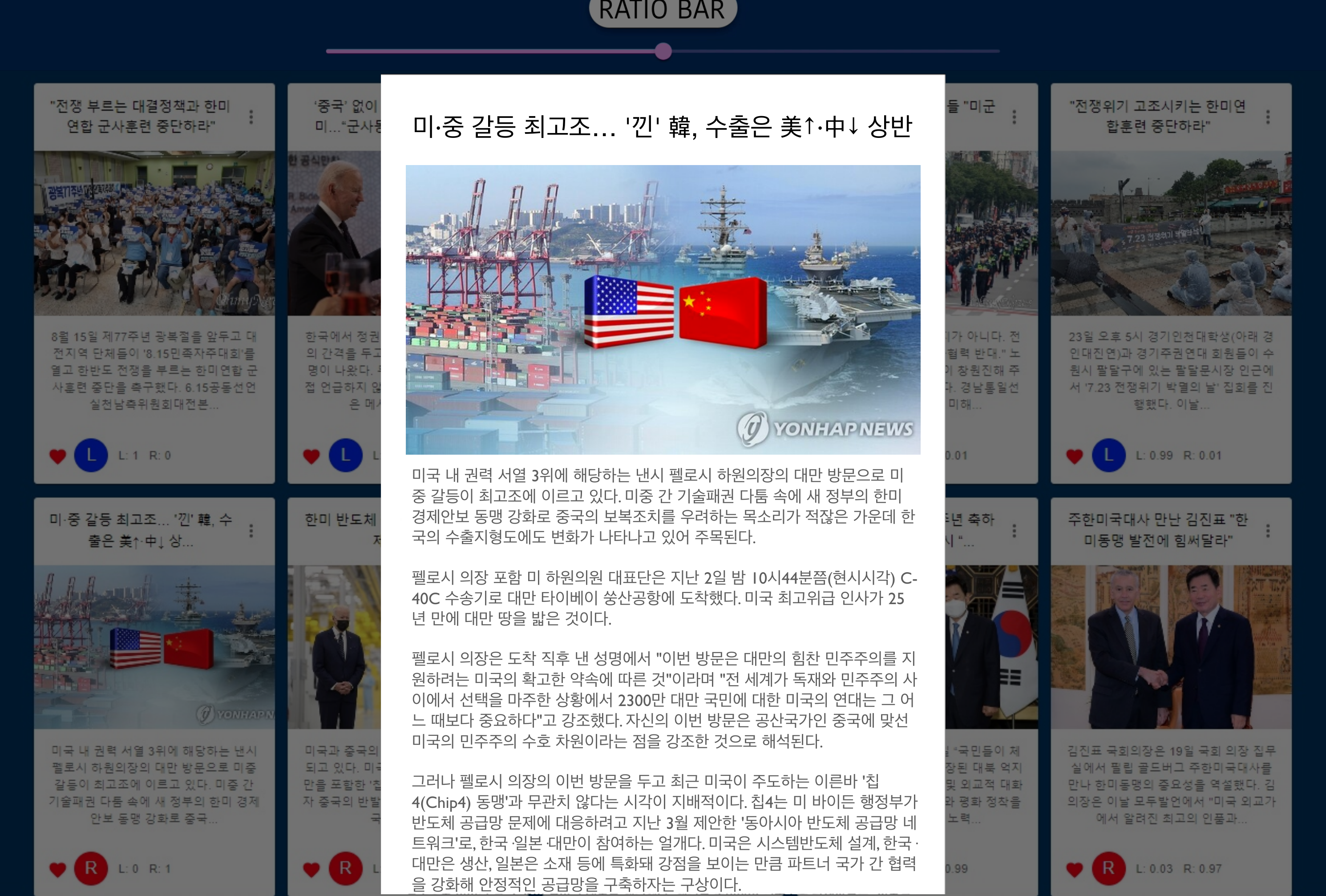}
  \vspace{-0.2cm}
  \caption{The visual example of News viewer (Visualization 1). This visualization presents the title-image-news content (summary) section in thumbnail format. When users click on some news, this visualization provides users with a full article.}
\vspace{-0.6cm}
  ~\label{fig:VE1-add}
\end{figure}

\subsection{Visualization 1: News viewer}\label{HH-1}
\textit{News viewer} provides users with 10 news articles related to specific political issues along with quantitative political-related information, such as political stance and level of political extremeness.
This visualization presents the title-image-news content (summary) section in thumbnail format. Similar to the reviewer's observation of real-world information consumption behavior, participants could get a rough idea of the news from the title and consume the full article by clicking on it (Figure~\ref{fig:VE1-add}). During the user study, we explained to the participants that they could access the full news article by clicking on the title, image, and other elements.
We used our model (Sec.~\ref{DvpAI}) for political stance and level of political extremeness prediction and applied it to \system. This visualization offers two functions---the first is a five-scale ratio bar (1: more conservative, 5: more liberal), allowing users to adjust the ratio of articles with a conservative political stance to those with a liberal stance on the basis of these users’ needs. It also offers a toggle button for a level of political extremeness ranging from `not extreme at all' or `very extreme', allowing users to adjust the extremeness of the articles' political stances (Sec.~\ref{barbutton}). With these functions, we expect that users will have opportunities to control and interact with the results recommended by the AI model based on their needs. We will explain our political prediction model and each function in more detail in the next sections.

\subsubsection{AI to predict a political stance}\label{DvpAI} 
For Visualization 1 (News Viewer, Sec.~\ref{HH-1}), we developed AI to predict the political stances of news articles. To accurately display news articles that have the correct political stance and level of political extremeness, we developed a graph-based political classification model. Through conducting user studies, we figure out five main factors (i.e., context, tone, frequently used words, person, and keyword) when identifying the political stance of a news article. We developed an AI model to predict the political stance of news articles by reflecting the results of our user study on building the model. The model consists of two key components: (1) hierarchical attention networks to learn the relationships among words/sentences in a news article with a three-level hierarchy (i.e., word-level, sentence-level, and title-level) and (2) knowledge encoding to incorporate both common and political knowledge for real-world entities, necessary for deeply understanding a news article, into the process of predicting the political stance of a news article. The model yielded a higher performance (92.26\% accuracy) than other existing state-of-the-art models. In the Appendix, we have included an explanation of AI-related details.

\begin{table}[h]
\centering
\caption{Paired policy dimension~\cite{lowe2011scaling} and the news themes of \system.}
\vspace{-3mm}
\label{tab:newstheme}
\setlength\tabcolsep{5pt}
\begin{tabular}{r||l}
\toprule
 
\textbf{Policy dimension} &\multicolumn{1}{c}{\textbf{New theme of \system}} \\

\midrule

\multirow{2}{*}{Labor policy}                & 1. Illegal strike of confederation of unions  \\ 
                                             & 2. Minimum wage increase  \\ 

\midrule

\multirow{2}{*}{Political corruption}        & 3. Impeachment of the Minister of Justice  \\ 
                                             & 4. Thesis plagiarism of the first lady  \\

\midrule

Human rights                                 & 5. Right of the disabled  \\ 

\midrule

Military                                     & 6. U.S. military alliance   \\ 

\bottomrule
\end{tabular}
\end{table}

\subsubsection{News information with political stance information.} 
We initially selected six topics from the ``headline news'' that dealt with important political issues in the authors' country (Table \ref{tab:newstheme}). Lowe~\cite{lowe2011scaling} revealed representative topics and frames shared by the conservative and liberal camps. One of the key considerations in our selection of news topics is the extent to which particular news articles receive public attention. Therefore, we sorted the news keywords according to their popularity within designated hot topic news categories on the news portals and then selected topics according to Lowe's policy dimension. We chose keywords related to the topics and crawled news articles from a news portal. We crawled a total of 600 news articles from major news media that represent either conservative or liberal views. We then used these news articles as the test dataset for our model and obtained the label for each news article. Finally, for each news topic, we prepared 20 news articles: 10 conservative and 10 liberal. The news articles for each stance consisted of five highly extreme (as a classification result, 95\% probability) and five moderately extreme (80\% probability) news articles.

\subsubsection{Ratio bar and extremeness button.}\label{barbutton}
\textit{News viewer} has two functions. The five-point-scale ratio bar allows users to adjust the ratio of news articles by political stance [10 conservative: 0 liberal, 7:3, 5:5, 3:7, 0:10]. In addition, by manipulating the level of an extremeness toggle button, users can sort the news in ascending/descending order by the value of extremeness. 
This enables users to consume news articles suggested by the AI model in a more interactive fashion, giving them more initiative. By clicking the thumbnail, users can check the body of the news article, which is displayed on a separate pop-up screen. They can also adjust the news ratio (Figure~\ref{fig:HereHear}, 1-A) and extremeness sorting (Figure~\ref{fig:HereHear}, 1-B).

In summary, \textit{News viewer} has the following order of usage. 
\begin{itemize}
  \vspace{-0.1cm}
    \item News articles with political stance information from AI help users quantitatively understand the idea of a certain news topic from different standpoints.
    \item The ratio bar allows users to easily adjust news articles by political stance.
    \item The extremeness button allows users to sort 10 news articles according to their level of political extremeness.
\end{itemize}

\begin{figure}
\centering
  \includegraphics[width=0.85\columnwidth]{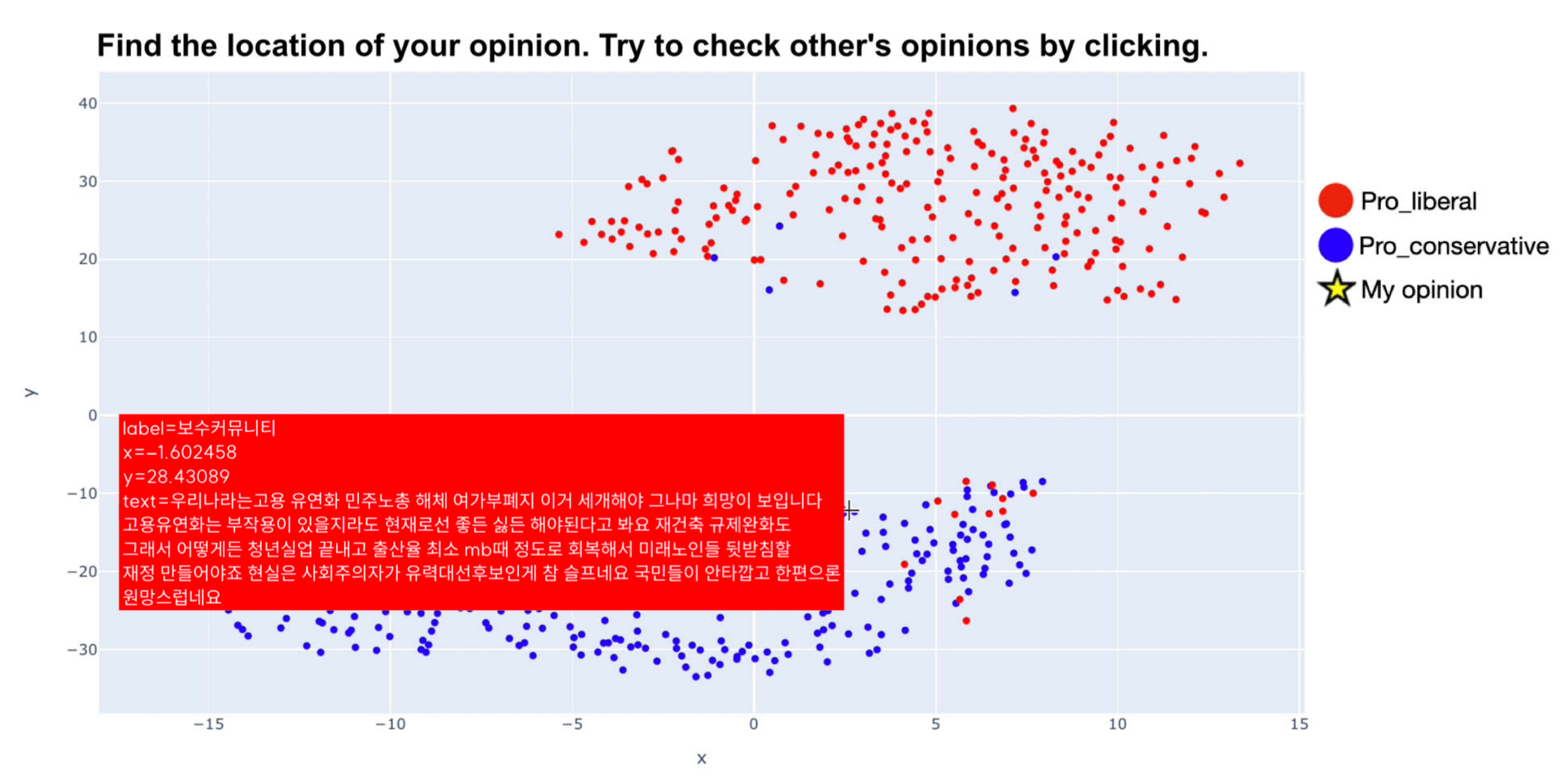}
  \vspace{-0.2cm}
  \caption{The visual example of My opinion (Visualization 2). This visualization presents the location of users' own opinions between pro-liberal and pro-conservative comments on a map interface. In addition, if users click comments, they can explore and examine the opinions of others.}
\vspace{-0.7cm}
  ~\label{fig:VE2-add}
\end{figure}

\subsection{Visualization 2: My opinion map}\label{HH-2}
In \textit{My opinion map}, users can submit their opinions about a specific political issue in the comment form (Sec.~\ref{MyThought}) and confirm the position of their own opinions between pro-liberal and pro-conservative comments presented in a map interface (Sec.~\ref{Mapping}). While many people leave comments and interact with others on online community platforms, they do not know how much their thoughts are related to other people's thoughts or how extreme their thoughts are in terms of political stance. The objective of this visualization is to give users opportunities to look back on their opinions by comparing them with others' opinions and accessing the opinions of different political stances. The user's comments will go through the natural language processing model, which generates the vector of the comment to be displayed on the map visualization. Here, the user has the initiative of leaving their comment, seeing where their comment is located, checking other users' comments, and possibly expanding their perspectives through the use of and interaction with \textit{My opinion map} (Figure~\ref{fig:VE2-add}). This visualization mainly consists of two functions as follows.

\subsubsection{My thought}\label{MyThought}
\textit{My thought} was designed to allow users to freely type in their own opinion about the news articles they saw in \textit{News viewer} (Figure~\ref{fig:HereHear}, 2-A). 

\system also provides 20 examples of comments so that users can choose one from the comment list. Those 20 comments were collected from online communities. We first collected 1,000 comments (500 per political stance) by topic. We then randomly sampled 50 comments per political stance. Two researchers checked whether those comments well represent each political stance. Finally, 10 comments were selected as example comments for \textit{My thought}.

\subsubsection{Opinion mapping}\label{Mapping}
\textit{Opinion mapping} was designed to vectorize the opinion of a user, giving the user's position in a two-dimensional space (Figure~\ref{fig:HereHear}, 2-B). 
To effectively embed the text data (i.e., comments), we used the NLP AI model [anonymized for review] trained with about 180 million sentences of news comments. This model is based on a transformer architecture and has shown great performance in extracting text features.

To construct opinion datasets on conservative and liberal stances, we collected additional sets of comments on six topics from two online communities [community names anomyized] that represent either conservative or liberal views. 

A total of 14,148 (Conservative: 7,074 and Liberal: 7,074) comments were collected. We divided these data into 75\% and 25\% as the training and test datasets, respectively. We then used the training dataset to fine-tune the NLP AI model and finally obtained a model that is suitable for identifying the political stance of the comment. Overall, the model achieved 94.46\% accuracy in multi-class classification tasks after 12 epoch fine-tuning runs.

For clustering and map visualization, we used t-SNE~\cite{van2008visualizing}. To obtain embedding representation for conservative and liberal comments, we extracted the 768-dimension vector of the special classification (CLS) token of the model by topic. When users submit their opinion on a topic, the CLS token of the sequence is extracted by inference through the same fine-tuned model. Then, the CLS token vector is reduced by a two-dimensional vector based on t-SNE, and \system visualizes that vector on the map along with the vectors of other comments from the online communities.

The red and blue plots of the visualization map (Figure~\ref{fig:HereHear}, 2-b) refer to the comments on the conservative and liberal stances, respectively. The yellow plot refers to the comment submitted by the user. In addition, the user can check the comment text by hovering their mouse over the dot on the plot. Through this process, users can check where their opinions (\textit{Here}) are located among the opinions of other users in their online communities. Through \textit{My opinion map}, we expect users to become more aware of various opinions of people from different political groups (\textit{Hear}) and expand their perspectives toward news information.

In summary, \textit{My opinion map} has the following order of usage. 
\begin{itemize}
    \item Through \textit{My thought}, users can write their thoughts on specific issues.
    \item Through \textit{Opinion mapping}, users can find out where their thoughts are located among the opinions of conservative and liberal communities and become more aware of people's different reactions and thoughts.
\end{itemize}

\begin{table}[h]
\centering
 \caption{Demographic information of the participants in user studies and in-depth interviews (total: 94 participants, 10 interviewees).}
\vspace{-3mm}
\label{tab:demo}
\setlength\tabcolsep{8pt}
\begin{tabular}{rr||cc|cc}
\toprule

\multicolumn{2}{c}{{\textbf{Demographic group}}}                                & \multicolumn{2}{c}{\textbf{User study}} & \multicolumn{2}{c}{\textbf{Interview}} \\ 
\multicolumn{2}{c}{}                                                                           & \textbf{N}     & \textbf{\%}    & \textbf{N}     & \textbf{\%}    \\ \hline

\midrule

\multirow{2}{*}{Gender}                  & Female       & 46 & 49\%  & 5 & 50\%  \\ 
                                         & Male         & 48 & 51\%  & 5 & 50\% \\

\midrule

\multirow{3}{*}{Age}                     & 19~29        & 39 & 41\%   & 4 & 40\% \\ 
                                         & 30~39        & 25 & 27\%   & 3 & 30\% \\
                                         & 40~49        & 30 & 32\%   & 3 & 30\% \\                        

\midrule

\multirow{3}{*}{Political Interest}      & High         & 17 & 18\%   & 4 & 40\% \\ 
                                         & Middle       & 48 & 51\%   & 3 & 30\%\\
                                         & Low          & 29 & 29\%   & 3 & 30\%\\  

\midrule

\multirow{3}{*}{Political Stance}        & Conservative & 19 & 20\%   & 3 & 30\%  \\ 
                                         & Moderate      & 57 & 61\%   & 4 & 40\%\\
                                         & Liberal      & 18 & 19\%   & 3 & 30\%\\  
 
\bottomrule
\end{tabular}
\end{table}

\section{User study}\label{US}
Our experiment had two research purposes. The first purpose was to see whether the participants would realize the importance of information diversity after using \system (RQ2). The second purpose was to examine whether the degree of acceptance of information having diverse perspectives after using \system differs from the user characteristics (RQ3). 

In evaluating ConsiderIt ~\cite{kriplean2012supporting}, the authors investigated how the participants explored diverse opinions from system use and how this exploration helped the participants form balanced opinions. In the study of StarryThoughts~\cite{kim2021starrythoughts}, the authors evaluated whether the system effectively supported users’ exploration of diverse opinions and how users’ opinions and attitudes towards the issue would be affected. Inspired by previous studies, we defined the following research questions to understand the effectiveness of \system in encouraging users to consider diverse perspectives when consuming news information and in making them aware of the importance of information diversity. Furthermore, we conducted interviews with participants for identifying the pros and cons of \system.

\subsection{Participants}

We recruited participants through a professional survey company. 
We originally recruited 122 individuals but excluded 28 who did not meet the requirements of the study (e.g., did not complete the survey or did not use \system every day). 

A total of 94 participants were considered for the data analysis. Table~\ref{tab:demo} shows the detailed deomgraphic information of participants. The participation criteria were as follows. 
First, the participants must be adults (at least 18 years old). Second, they must have the same nationality and live in the same country as the authors (anonymized). Third, they must regularly consume online news (more than five times a week) from news portals and online communities. The study was approved by the institutional review board (IRB) of the author’s institution. The participants were given the web link of \system. Once they reached the site, they were presented with an informed consent page (explaining the study goal, procedure, duration, compensation, possible benefits/risks, information protection, opt-out option, and contact information). Only the participants who agreed to the study could start the survey. Furthermore, we conducted in-depth interviews with ten participants who were randomly selected among the 94 participants, considering the distribution of demographic information.

\vspace{-2mm}
\subsection{Procedures}
The participants were asked to complete the pre-survey, use \system once a day for about 15 minutes (for three days), and complete the post-survey. They read news articles about two topics per day. Thus, the participants accessed and interacted with a total of six topics during the study.

Only those who completed the pre-survey, used \system for three days (at least three times a day), and completed the post-survey, were given \$17.5 after the study. For improving the quality of our analysis, participants who did not use \system submit their opinions were disqualified and thus removed from the dataset for analysis.

\subsubsection{Demographic questions}
In the pre-survey, the participants were asked to provide demographic information, including gender (male, female, and others), age (19-29, 30-39, and 40-49), political interest (measured on a five-point Likert scale where 1 is ``not at all'' and 5 is ``very interested''), political stance (measured on a five-point Likert scale where 1 is ``very liberal'' and 5 is ``very conservative''), and media usage (measured on a five-point Likert scale where 1 is ``never'' and 5 is ``very often''). 

Specifically, in the case of political stance, we categorized individuals who selected 1 or 2 as liberal, 3 as moderate, and 4 or 5 as conservative. In the case of political interest, we categorized individuals who selected 1 or 2 as a low interest group, 3 as a middle interest group, and 4 or 5 as a high interest group.

\begin{table*}[]
\vspace{-0.3cm}
\centering
\caption{The EC breaking questions~\cite{dubois2018echo} based on the features of an echo chamber effect (Disagree, Different, Confirm, Offline, and Changed). In the pre-survey, we asked about the participants’ current behaviors. In the post-survey, we asked about their willingness to adopt behaviors that could prevent the echo chamber phenomenon.}
\scriptsize
    \small
    \resizebox{\textwidth}{!}{
    \begin{tabular}{ p{2cm} p{7cm} p{7cm} }
    \toprule
    \multicolumn{1}{c}{\textbf{Question}} & \multicolumn{1}{c}{\textbf{Pre-survey}} & \multicolumn{1}{c}{\textbf{Post-survey}} \\ \midrule
    \multicolumn{1}{c}{Q1} & 
    How often do you read something you \textbf{DISAGREE} with? &
    How often will you read something you \textbf{DISAGREE} with?
    \\
    \multicolumn{1}{c}{Q2} & 
    Have you ever checked a news source that is \textbf{DIFFERENT} from what you normally read? &
    Will you check a news source that is \textbf{DIFFERENT} from what you normally read?
    \\
    \multicolumn{1}{c}{Q3} & 
    Do you try to \textbf{CONFIRM} information you find by searching online for another source? &
    Will you try to \textbf{CONFIRM} information you find by searching online for another source?
    \\
    \multicolumn{1}{c}{Q4} & 
    Do you try to confirm information by checking a major \textbf{OFFLINE} news medium? &
    Will you try to confirm information by checking a major \textbf{OFFLINE} news medium?
    \\
    \multicolumn{1}{c}{Q5} & 
    Thinking about recent searches you have performed online using a search engine, how often have you discovered something that \textbf{CHANGED} your opinion on an issue? &
    How often will you discover something that \textbf{CHANGES} your opinion on an issue?
    \\
    \bottomrule
    \end{tabular} }

    \label{tab:ECQuestion}
    \vspace{-0.5cm}
\end{table*}
\subsubsection{Echo chamber breaking score (EC breaking score)}
In both the pre- and post-surveys, we asked the EC breaking questions to measure the importance of diverse perspectives in information consumption (RQ2 and RQ3). Table~\ref{tab:ECQuestion} outlines the specific survey questions we used in the pre- and post-surveys. We used a five-point Likert scale for all the questions (1 is ``Almost never'' and 5 is ``Nearly always.'') To measure the effects of mitigating echo chambers and promoting information diversity, we selected the EC breaking score questions based on previous research~\cite{dubois2018echo}. The questionnaire developed in~\cite{dubois2018echo} uses five dependent variables (i.e., Disagree, Different, Confirm, Offline, and Changed), which collectively assess different aspects of echo chambers. These variables gauge the frequency with which users encounter opposing opinions or information, the extent to which users take action to disengage from echo chambers, and the extent to which users expose themselves to unfamiliar or new/different information. According to Dubois et al.~\cite{dubois2018echo}, these variables can be used collectively to measure an individual's echo chamber. Therefore, recognizing the importance of diverse information and information sources is closely related to mitigating echo chambers. Jeon et al.~\cite{jeon2021chamberbreaker} used the EC breaking score to evaluate whether their proposed system helped participants realize the importance of diverse perspectives in consuming information.

\begin{figure}
\centering
  \vspace{-0.5cm}
  \includegraphics[width=1\columnwidth]{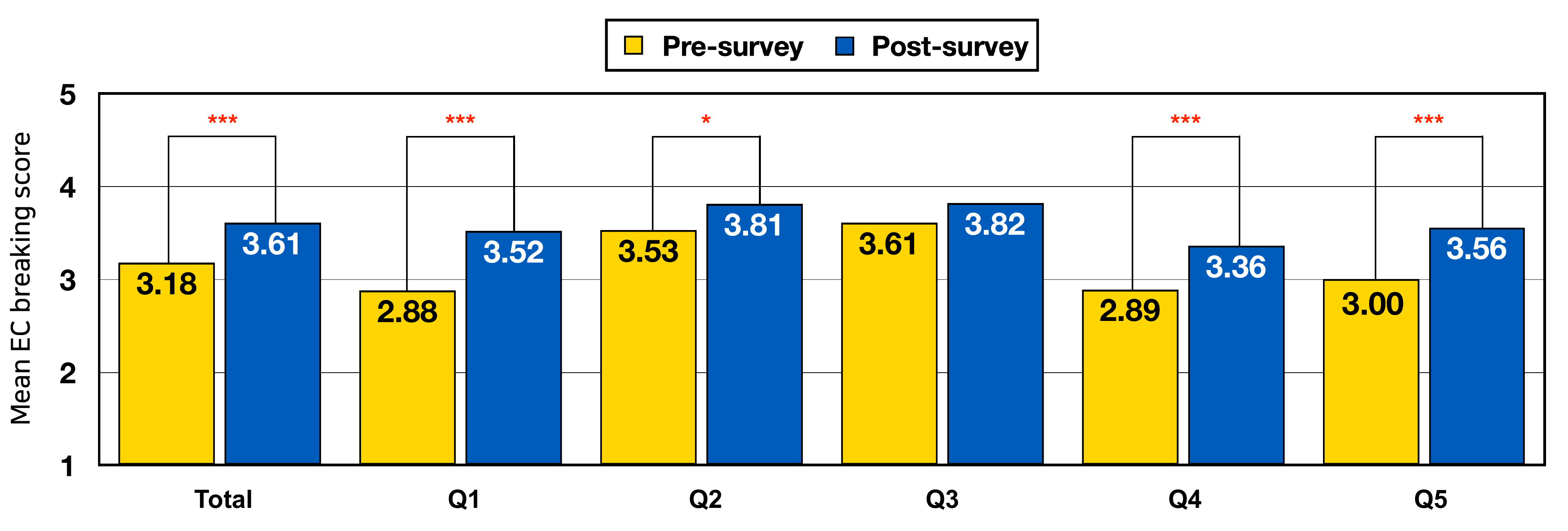}
    \vspace{-0.7cm}
  \caption{Bar plots showing the differences in the EC breaking scores between the pre- and post-surveys (***\textit{p}<.00016, **\textit{p}<.00166, *\textit{p}<.00833). All groups showed significant differences between the pre- and post-surveys.}

  \label{fig:results_ec_breaking_score_all}
\end{figure}
\section{Results}

Given that the participants were volunteers and did not represent any particular population, the general distribution of the sample was somewhat skewed toward the younger generation (19-29, 41\%), having reasonable political interest (51\%), and having a neutral political stance (61\%).
Table~\ref{tab:demo} summarizes the demographic information of our study participants.

To answer RQ2, we examined the impacts of \system on the recognition of diverse perspectives in consuming information, as measured by the EC breaking score. We analyzed the difference in the EC breaking scores between the pre- and post-surveys from all participants (Sec.~\ref{ECB}).

To answer RQ3, we conducted in-depth analyses to explore the influence of demographic factors on the EC breaking score (Sec.~\ref{ECBbyDemo}). This analysis was motivated by prior research which showed the influence of various demographic factors, such as age or political stance, on shaping one's attitudes toward news consumption~\cite{napier2008conservatives,furlong2006young}. For the statistical analysis, we used an open source tool, jamovi~\footnote{https://www.jamovi.org/}.

\subsection{EC breaking score}\label{ECB}
A paired sample t-test was conducted to measure whether \system helped the participant aware of the importance of having diverse perspectives in news consumption.
For multiple comparisons, we applied the Bonferroni correction~\cite{bland1995multiple} to counteract the multiple comparisons problem. This problem arises when performing a large number of statistical tests in the same experiment, since the more tests we perform, the higher the probability of obtaining at least one statistically significant test. 
The adjusted p-values were as follows: ***p<.00016 for a significance level of 0.001, **p<.00166 for 0.01, and *p<.00833 for 0.05.
As a result, the mean of the EC breaking scores between the pre- and the post-surveys showed a significant difference (\textit{t}(94)=-8.54, \textit{p}<.00016). In addition, the significant result was shown in each of the five EC breaking questions as follows: Q1: \textit{t}(94)=-5.86, \textit{p}<.00016, Q2: \textit{t}(94)=-3.16, \textit{p}=.0021, Q3: \textit{t}(94)=-2.05, \textit{p}=.04272, Q4: \textit{t}(94)=-4.74, \textit{p}<.00016, Q5: \textit{t}(94)=-5.05, \textit{p}<.00016. Figure~\ref{fig:results_ec_breaking_score_all} illustrates these results.

\begin{figure}
\centering
  \includegraphics[width=1\columnwidth]{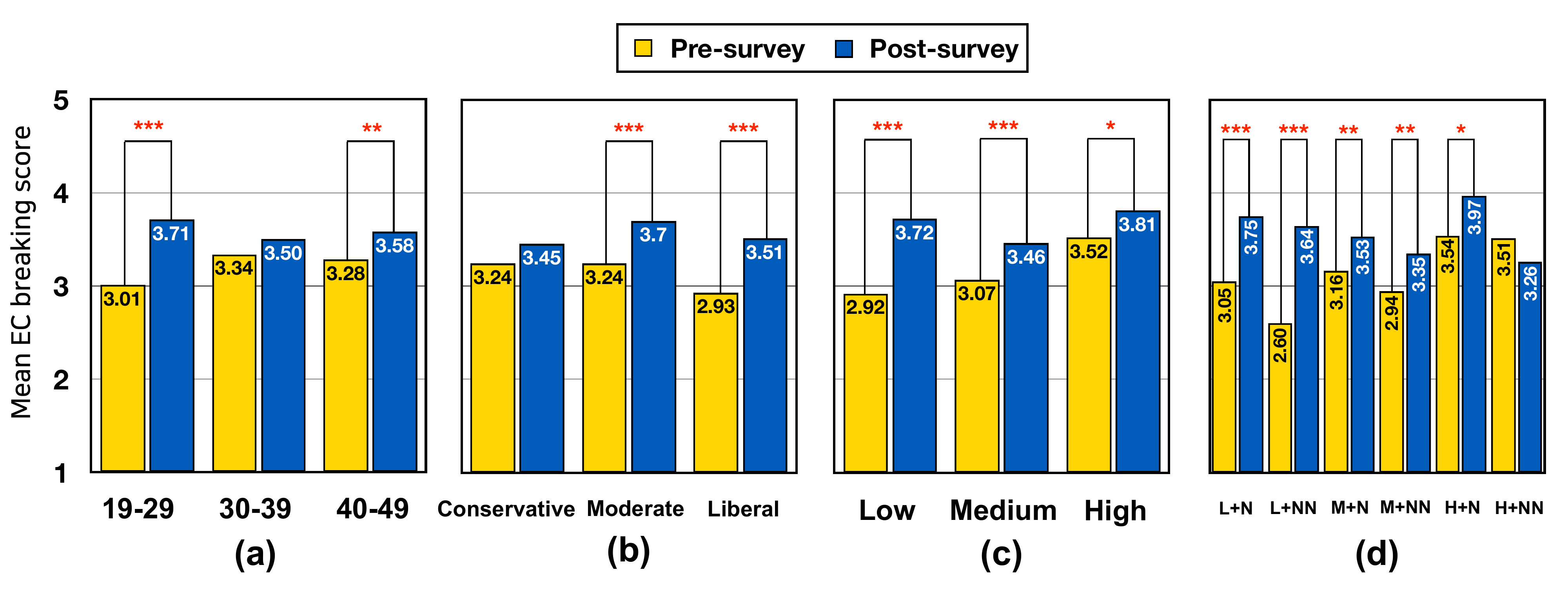}
  \vspace{-0.4cm}
  \caption{Bar plots showing the differences in the EC breaking scores between the pre- and post-surveys according to (a) age (***\textit{p}<.00025, **\textit{p}<.0025, *\textit{p}<.0125), (b) political stance (PS) (***\textit{p}<.00025, **\textit{p}<.0025, *\textit{p}<.0125), (c) political interest (PI) (***\textit{p}<.00025, **\textit{p}<.0025, *\textit{p}<.0125), (d) PI$\times$PS (L+N: Low+Neutral, L+NN: Low+Non-neutral, M+N: Middle+Neutral, M+NN: Middle+Non-neutral, H+N: High+Neutral, H+NN: High+Non-neutral) (***\textit{p}<.00014, **\textit{p}<.00142, *\textit{p}<.00714).}

  \label{fig:results_ec_breaking_score_by_demographic}
\end{figure}

While it is noticeable to see significant increases in all questions, we paid particular attention to the results of Q1 and Q2. Q1 relates to the consumption of reading materials (e.g., news), which aligns with the purpose of \textit{News viewer}. Q2 relates to seeking out additional information from additional/alternative sources online (e.g., other people's opinions in online communities), which aligns with the purpose of \textit{My opinion map}. The increases in the scores for Q1 and Q2 highlight the effective design rationale of \system, as it successfully facilitated participants' recognition of the importance of embracing diverse perspectives when consuming information. We will discuss this aspect in more detail in Section~\ref{discussion}.

However, we found that the result for Q3 was not significant. This result is primarily attributed to the generally high score for Q3 compared to the other questions in both the pre- and the post-tests (Figure~\ref{fig:results_ec_breaking_score_all}). Despite this non-significant result, it is important to recognize that the the scores for all questions increased after using \system. This overall positive trend in the scores still reinforces the main contributions of our work and the effectiveness of \system in promoting the recognition of diverse perspectives in information consumption. We found similar results in a previous study by Jeon et al.~\cite{jeon2021chamberbreaker}. They developed a game interface aimed to mitigate echo chambers and evaluated its effectiveness using the same EC breaking score. In their findings, Q1 and Q3 did not show statistical significance. Nevertheless, the authors highlighted the statistically significant differences observed in Q2, Q4, and Q5, as well as the overall increase in scores and mean values.

\subsection{EC breaking score by demographic}\label{ECBbyDemo}
Prior studies~\cite{boxell2017internet,napier2008conservatives} presented the impacts of demographic factors on changing people’s attitudes toward information consumption.
Figure~\ref{fig:results_ec_breaking_score_by_demographic} illustrates the differences in the EC breaking scores based on age, gender, political stance, political interest, and the interaction effect between a political stance and political interest. 

A two (pre- and post-survey) $\times$ n (the number of features in each demographic group) mixed ANOVA~\cite{howell2012statistical} was conducted to measure whether a particular demographic group shows a significant impact on the results of the EC breaking score. The reason to use mixed ANOVA is to compare the mean differences between groups that have been split into two factors, where one factor is a within-subjects factor (pre- and post-surveys) and the other factor is a between-subjects factor (demographic groups). The post-hoc comparisons were also conducted with the Bonferroni correction~\cite{bland1995multiple}. 

First, for age, we identified significant differences in the EC breaking scores among the age groups between the pre- and post-surveys (\textit{F}(2, 91)=13.5, \textit{p}<.{00025). As shown in Figure~\ref{fig:results_ec_breaking_score_by_demographic}-a, the post-hoc test showed that the difference was evident for the 20 age group (\textit{p}<.00025) and for the 40 age group (\textit{p}=.00028).

Second, for political interest, we identified significant differences in the EC breaking scores among the political interest groups between the pre- and post-surveys (\textit{F}(2, 91)=7.02, \textit{p}=.001). As shown in Figure~\ref{fig:results_ec_breaking_score_by_demographic}-b, the post-hoc test showed that the difference was evident for all factors: the high political interest group (\textit{p}=.01062), the middle political interest group (\textit{p}<.{00025), and the low political interest group (\textit{p}<.{00025).

Third, for a political stance, we identified significant differences in the EC breaking scores among the political stance groups between the pre- and post-surveys (\textit{F}(2, 91)=2.94, \textit{p}=.058). As shown in Figure~\ref{fig:results_ec_breaking_score_by_demographic}-c, the post-hoc test showed that the difference was evident for the liberal group (\textit{p}<.00025}), the neutral group (\textit{p}<.00025).

Lastly, to examine the influence of politically-related factors, we measured the interaction effect between the political interest (high vs. middle vs. low) and political stance (neutral vs. non-neutral (conservative and liberal)) on the EC breaking score. 

As a result, we identified significant differences in the EC breaking scores among the groups between the pre- and post-surveys (\textit{F}(5, 88)=4.06, \textit{p}=.002). The post-hoc test showed that the difference was evident for all cases, except for the high and non-neutral groups (\textit{p}=.39493).

\subsection{Log data for opposing viewpoints}
Our user study results show that the UI of \system had a positive impact on facilitating the consumption of news article bodies and exposure to opposing viewpoints. Throughout the three-day user study, participants engaged with \system by consuming their usual news content. Notably, a significant number of participants were observed to read the entire news articles provided. With a daily allocation of 20 news articles (10 conservative and 10 liberal), we confirmed through log analysis of scrolling behavior that participants read an average of 7.3 articles (SD: 4.37) during the study period.

Interestingly, we found that participants showed a significant consumption rate of news articles that present opposing political stances. The consumption ratio between news articles supporting their preferred political stance (PS) and those with an opposing PS was 0.48:0.52. This indicates that our study participants actively read news articles from both sides of the political spectrum. We believe that these findings hold particular significance when we look back at the responses of six participants in the formative study (Sec.~\ref{Formative}), who reported that they rarely consumed news articles that presented opposing PS. In addition, similar trends were observed based on participants' political ideology (PI) and PS. The conservative participants had a consumption ratio of 0.42:0.58 for news articles opposing their PS, while the liberal participants had a ratio of 0.56:0.44. The participants with a high PI had a consumption ratio of 0.45:0.55 for news articles opposing their PS, whereas those with a low PI exhibited a ratio of 0.55:0.45.

These findings highlight the meaningful impact of \system in promoting the consumption of news articles that present opposing viewpoints. They also indicate a shift from participants' previous habits of limited exposure to news articles with opposing political stances to some extent. The observed trends based on participants' political ideology and preferred political stance further support the efficacy of \system in encouraging a more balanced consumption of news information.

\subsection{Interview}\label{interview}

During the interview, the participants explained in detail how AI helps users consume news information with quantitative aids (Visualization 1-\textit{News viewer}) and compare other opinions based on self-awareness (Visualization 2-\textit{My opinion map}). When reporting interview quotes, we use $P^{X}_{u}$  to denote participant number \textit{X} in the user study. After the interviews were completed, we applied thematic analysis and iterative open coding~\cite{smith2015qualitative} to analyze the interview transcripts. Two researchers who specialized in qualitative analysis coded and analyzed the transcripts for emerging themes and the findings were discussed among the co-authors of this paper iteratively until a consensus was reached.

\subsubsection{Mitigating echo chamber effects with \system}
All participants answered that \system provided sufficient support for having a balanced perspective in consuming information. They mentioned that \system has high efficiency in information consumption by providing two functions that deal with the two phases of consuming political information in the real world. Participants mentioned that 
watching news pertaining to a specific political event and comparing their thoughts with other opinions in an interface can effectively increase the probability of exposure to diverse perspectives. For example: \textit{``The feature of providing preferred news by portal sites injects biased political ideology, so I think this study showed one possible way to solve these side effects''} ($P^{5}_{u}$). \textit{``If there is a site that provides these functions, consume news, it might be useful''} ($P^{3}_{u}$).

\textit{News viewer} provides 10 news articles related to specific political issues along with quantitative political-related information, such as political stance and level of political extremeness as objective standards in information consumption. Participants mentioned \textit{News viewer} was useful to consume information with quantitative political stances and extremeness. For example: \textit{``It was convenient to use this function to view progressive/conservative news articles at once and understand their characteristics''} ($P^{2}_{u}$). \textit{``It is helpful to obtain the most objective information corresponding to the midpoint while viewing articles from various perspectives''} ($P^{4}_{u}$). In addition, participants responded that \textit{News viewer} was also helpful in navigating diverse news articles based on the standard. For example: \textit{``Originally, it was necessary to visit several portal sites to check articles of a specific political orientation, but this function made it possible to efficiently search for this news myself''} ($P^{1}_{u}$). \textit{``It was helpful to read articles from different viewpoints and check the correctness of my thoughts by moving the news balance bar''} ($P^{6}_{u}$).

\textit{My opinion map} allows users to submit their opinions about a specific political issue in the comment form and to identify the position of their own opinions between pro-liberal and pro-conservative comments presented in a map interface. Participants mentioned that while many people leave comments and interact with others on online community platforms, they do not know how diverse other people think and how extreme their thoughts are. \textit{My opinion map} helped users easily grasp the enormous amount of others' thoughts related to a particular political event. For example: \textit{``It was beneficial to see the biased opinions around me and selectively check others' opinions''} ($P^{5}_{u}$). \textit{``The function to visualize various opinions was impressive, and it was good in terms of diversity''} ($P^{2}_{u}$). In addition, participants mentioned that \textit{My opinion map} can be also helpful to the participants who do not have strong beliefs about politics. For example: \textit{``Political beginners are likely to fall into political polarization by following only a few news articles without knowing their stance, so this function seems to help mitigate it''} ($P^{8}_{u}$). \textit{``I am not that interested in political stances in news articles, but being able to see what kind of political stance I had through the map seems quite useful''} ($P^{9}_{u}$).
In addition, \textit{My thought} with pre-written examples was to help users formulate their thoughts. During the three-day study period, all participants were instructed to submit their opinions on a daily basis. Based on our log data, we observed that approximately 90\% of the participants opted to select and utilize the provided examples rather than submitting entirely personal thoughts. Through interviews, we found the importance of pre-written statements as anchors to guide their thought processes. For example, \textit{some participants stated that ``Examining the content of the statement proved beneficial. Although certain statements could be employed as is, I personally prefer crafting my own.''} ($P^{8}_{u}$). \textit{``The opinions provided through the ‘My Thoughts’ feature were greatly helpful in organizing my thoughts into writing. Over the course of three days, I utilized those suggestions as a foundation and supplemented them with my own views to use the \system features.''} ($P^{3}_{u}$).

\subsubsection{Backfire effect and Political interest}
One of the critical aspects of \system is the potential resistance that individuals may experience when exposed to different perspectives. We asked the participants about their concerns or reservations when accessing the AI results. First, in the context of \textit{News viewer}, participants expressed difficulty in comprehending news articles presenting an opposing political stance, which sometimes made them uncomfortable. For example: \textit{``I didn't want to see news articles with the opposite political stance from mine because it was difficult to understand''} ($P^{7}_{u}$). \textit{``I was uncomfortable because it was from the news media that I usually thought negatively about.''} ($P^{9}_{u}$).
Second, in the case of \textit{My opinion map}, participants reported being repulsed by strong expressions, including the use of swear words, when examining opinions from different standpoints. For example: \textit{``I felt rejected because there were many unreasonable opinions and radical expressions about a certain issue''} ($P^{10}_{u}$).

Our interview results highlight an interesting aspect regarding the level of tolerance towards exposure to opposing perspectives, which varied based on participants' political interests. First, the participants with relatively low political interest displayed a higher degree of flexibility in accepting the information with opposing perspectives in \textit{News viewer} and \textit{My opinion map}. They mentioned that exposure to different perspectives was a valid means of making informed decisions on political matters. Especially, the participants were satisfied with the ease of accessing diverse opinions with just a few clicks. $P^{6}_{u}$ respond that \textit{``I was able to check the fact that my opinions and political stance changed depending on the news article, and through the map function, I was able to find out that my political stance was neutral after all''}. $P^{8}_{u}$ mentioned that \textit{``As I read comments that have different perspectives, I realized that many opinions were different from what I think is correct''}.

Second, on the other hand, participants with relatively high political interests showed a greater resistance towards consuming information with conflicting perspectives. While they agreed with the importance of consuming information from balanced standpoints, they still found it difficult to accept ideas or perspectives contrary to their own on certain political issues. They felt that interfaces like \system somewhat forced them to read news articles or comments with which they might disagree. In particular, the participants tended to perceive themselves as being exposed to opposing opinions rather than diverse perspectives. $P^{4}_{u}$ mentioned that \textit{``I found it frustrating when anonymous comments insisted on radical viewpoints without any supporting evidence''}. $P^{7}_{u}$ respond that \textit{``I was embarrassed and uncomfortable because there were many opinions different from what I thought was right''}.

An important finding in our study was that \system played a little role for participants with strong political preferences and high interests, as their EC breaking scores decreased in the post-survey (Figure~\ref{fig:results_ec_breaking_score_by_demographic}-d). While most participant groups positively evaluated the role of \system and showed significant, positive changes in their EC breaking score, the contrasting result in this particular group may suggest the need for a different approach to delivering the positive effects of awareness of different standpoints when consuming news. Because the group already held firm political beliefs and had established information consumption methods, it may be natural for them to show resistance to the new method of information consumption. Since the purpose of \system is not to alter individuals' behaviors but to broaden their acceptance of various perspectives, this finding highlights the importance of employing a more thoughtful and tailored approach for a specific group.

\subsubsection{Need for having initiative and political interest}
Participants expressed their preferences and needs when using \system. They emphasized the importance of having diverse and segmented options that actively reflect their thoughts and allow them to interact with AI-based results. The needs of participants depend on visualization. In \textit{News viewer}, participants suggested further subdividing the scope and level of political stances. $P^{4}_{u}$) mentioned that \textit{``I think it was very useful to provide the political stance of the news quantitatively and allow me to control it according to my standards. However, it would be better if I had more flexibility in controlling the results of AI''} (. In \textit{My opinion map}, participants suggested having data from a wider range of online social communities, since the political identity can be different by communities. $P^{8}_{u}$ respond that \textit{``I think \textit{My opinion map} is really convenient because I could read both news articles and opinions. Especially, it was really interesting to check the location of my thoughts, which seemed to work well. However, it would be helpful if I could take the initiative in choosing online communities on \textit{My opinion map}. I don't want to follow the results of only two communities.''}. In summary, participants expressed a desire for more agency in their interaction with the AI in order to enhance the effectiveness of using \system and the applicability of AI in the real world.

Our interview results highlight an interesting aspect. The purpose of participants taking initiative differed depending on their level of political interest. First, the reason why participants with low political interest wanted to take the initiative was to promote balanced information consumption. They mentioned that taking the initiative and consuming a large amount of information, guided by the objectivity provided by AI, would lead to a more balanced perspective. $P^{2}_{u}$ mentioned that \textit{``I want to receive more news than what is currently available in \textit{News viewer}. If I can actively select and read the news that revolves around the topics I consider important, I believe I will be able to think more objectively.''}. $P^{3}_{u}$ respond that \textit{``While I have a fair amount of trust in AI-labeled perspectives \textit{News viewer}, it would be even better if my own criteria were included. This is because if I receive AI-generated perspective results that include my criteria, I can focus more on opposing opinions.''}. $P^{5}_{u}$ said that \textit{``I would like my community to be added to the function \textit{My opinion map}. If that happens, I believe I will be able to more intuitively grasp the differences between my own thoughts and opposing views.''}. In summary, the participants with lower political interest wanted the initiative to align with the original goal of \system, which may be the reason behind their preference.

Second, on the other hand, the reason participants with high political interest wanted to take the initiative was to consume the information that aligned with their own beliefs. While they understood the importance of consuming the news from different perspectives, their priority was accessing news and opinions that resonated with their own thoughts. $P^{4}_{u}$ mentioned that \textit{``I really liked the feature that allows the AI to adjust the political stance score in \textit{News viewer}. However, the news labeled as 100\% conservative by the AI seemed less conservative according to my standards. Having my criteria reflected in the AI`s score would help me find the conservative news I want.''}. $P^{7}_{u}$ respond that\textit{``In \textit{My opinion map}, I wish I could set the number of opinion samples. I would like to see more progressive opinions similar to mine.''}. In short, the reason why the participants with higher political interests wanted the initiative may be to strengthen their own perspectives, which diverges significantly from the original purpose of \system.

\section{Discussion}\label{discussion}
We have outlined our research methodology and introduced \system by incorporating insights of news consumers for the applicability of an AI-based web system in order to mitigate echo chambers (RQ1). Through the user study, we demonstrated the feasibility of \system. The results indicated that by using \system, the participants showed an increased awareness of the importance of information diversity (RQ2) and showed that the impact of \system was different by users' political interest (RQ3). We summarize the findings of the study and discuss their implications. We also report the limitations of the study and our plans for future work.

\subsection{Interface to mitigate echo chambers}
In this research, we first identified two primary design rationales to facilitate news consumption from diverse political standpoints: (1) providing political stance information of news articles by using AI, (2) identifying users' thoughts on political issues for comparing other opinions, and (3) taking into account the news consumption process, which consists of information understanding and opinion comparison/confirmation. 
Supporting (1) and (2) pertains to providing quantitative information and (3) pertains to moving one step forward to consuming information with diverse perspectives.  
Second, to operationalize these design rationales, we introduced \system, a system that leverages an AI model to predict the political stance of news articles. \system offers two visualization interfaces that aim to mitigate political polarization in political news consumption. 
Lastly, we presented the significant impacts of \system on increasing one's awareness of the importance of news consumption from diverse standpoints through the user study with 94 participants.

Both the quantitative and qualitative results of our user study confirmed that \system has an impact on mitigating political polarization and broadening one's perspectives on news consumption.}
We carefully designed and developed \system to support balanced information consumption by reflecting the key processes of online information consumption with respect to digital literacy. We also demonstrated the design rationales of \system through the user study in the wild and confirmed the effectiveness of our approach for the design of a computer-based tool for news consumption. 

\subsection{Design implications}
Our study results have shed light on the application of AI in mitigating echo chambers, considering the nature of tasks and human evaluations. This means that in an interface to mitigate echo chambers, it is necessary to carefully consider design implications to enhance usability based on users' characteristics. In the subsequent subsections, we will delve into the implications of these findings and explore strategies for effectively using AI results in the context of mitigating echo chambers. 

\subsubsection{Backfire effect based on political interest}

Social science studies have well documented the presence of the backfire effect~\cite{nyhan2014effective, taber2006motivated} in information consumption, where individuals tend to reinforce their existing beliefs and polarization when exposed to opposing viewpoints. Motivated reasoning leads individuals to selectively accept or reject information based on their preexisting beliefs and preferences, often disregarding the accuracy of the information~\cite{nyhan2010corrections, taber2006motivated}. This effect highlights the challenges of providing users with diverse views without a careful approach, as it may inadvertently strengthen polarization and reinforce existing beliefs.

Despite the potential side effects, research within the Human-Computer Interaction (HCI) community emphasizes the significance of offering diverse perspectives to facilitate balanced information consumption. Our findings, which demonstrate the variability of the backfire effect depending on the level of political interest of the user, suggest the need for different approaches tailored to individual political interests. When political interest is low, the advantages of promoting diversity outweigh the associated side effects, validating the current direction. On the other hand, when political interest is high, it may become crucial to proactively prioritize raising awareness of the importance of balanced consumption to mitigate the risk of echo chambers and associated side effects.

Examples of such efforts are the development of educational games to address information-related challenges. ChamberBreaker (CB)~\cite{jeon2021chamberbreaker} is a game interface designed to increase players' awareness of and ability to proactively respond to the echo chamber effect. By assuming the role of anonymous social media users, players engaged in activities that contribute to the formation of an echo chamber within their online community, exposing them to various scenarios highlighting the problems associated with echo chambers. Similarly, Fakenews game~\cite{roozenbeek2019fake} aims to preemptively expose players to the techniques used in creating misinformation, helping them develop cognitive immunity against real instances of fake or misinformation. 

However, taking into account the insufficient emphasis on the importance of political stance in previous studies, it is important to note that careful consideration and planning should be given to incorporating users' political interests and stances into interface design.

\subsubsection{More initiative, but different purpose based on political interest}

Previous research has identified a positive relationship between usability improvement and user initiative~\cite{horvitz1999principles, shneiderman1997direct, xiao2008mixed} in the development of a computer interface, which is one of the key aspects related to usability. Giving users a degree of control and initiative in interacting with computer services or functionalities has been shown to enhance user experience. In addition, as more services will be supported by AI capabilities, it will become increasingly important to give users the ability to take initiative in their interactions with AI to achieve a satisfactory user experience. Indeed, researchers in the field of human-AI interaction have emphasized the significance of user initiative in improving the usability of AI-based interfaces~\cite{oh2018lead, deterding2017mixed, koch2020imagesense, xiao2008mixed, yannakakis2014mixed}. For example, Oh et al.~\cite{oh2018lead} presented the importance of having initiative in improving usability in their interface, DuetDraw, which allows users to draw pictures with AI. Salvador et al.~\cite{salvador1996denver} discussed the timing of initiative (spontaneous and pre-planned initiatives) in their framework to maximize usability. While the importance of user initiative has been recognized, its application in supporting online news consumption while considering a reader's political stance and diverse perspectives remains unexplored. Therefore, in human-AI interaction, determining the scope and degree of user initiative is one of the important factors in computer interface design~\cite{rezwana2022designing}.

Our research findings align with those of previous studies and highlight the significance of user initiative, particularly in the context of political news consumption. Notably, the purpose of user initiative may vary based on one's level of political interest. Thus, the level of user initiative should be adjusted according to their political engagement. For users with higher levels of political interest, the initiative should aim to provide information that aligns with their existing beliefs. On the other hand, for users with lower levels of political interest, the goal is to allow them to consume a broader range of opinions in an efficient manner. Giving users too much initiative may lead to outcomes that contradict the intended purpose of the interface. Therefore, future interface designs for promoting balanced political news consumption should take into account users' levels of political engagement and carefully determine the appropriate level of user initiative. Effectively calibrating this factor is essential to foster diverse political discourse, mitigate the risk of echo chambers, and ensure the representation of a wide range of views.

\subsection{Limitations and future work}
We acknowledge the limitations of our study, which we plan to address in future work. We recognize that \system's content may not fully represent online news environments and that certain sensitive topics were excluded. To address this, we plan to expand \system with more diverse news topics and conduct additional user studies with broader demographic representation. While thumbnail and full news consumption patterns are crucial, we did not collect thumbnail interaction logs in this study, but we intend to explore these patterns in future research. To mitigate concerns of potential overestimation, we will incorporate additional log data on the consumption of opposing views in our study's evaluation phase. We also aim to investigate the trade-off relationship between diversity and potential side effects, such as the backfire effect.  Furthermore, we have identified the potential of having initiative in enhancing the usability of AI-based interface; thus, we intend to conduct a study on information consumption initiative by applying initiative-related research in other domains, such as the timing of initiative~\cite{salvador1996denver, alam2013computer} and specific tasks in which users want more initiative~\cite{jeon2021fashionq, oh2018lead}}

We believe "My opinion map" (Visualization 2) can be more interactive to improve usability, considering participant feedback. Potential future directions include integrating demographic information and filtering options for enhanced usability, as demonstrated in StarryThoughts~\cite{kim2021starrythoughts}. Identity integration can provide context for opinions, aiding user understanding of diverse viewpoints~\cite{dahlberg2001computer} and credibility evaluation, with the potential to reduce stereotypes and foster democratic discourse~\cite{brauer2011increasing}. Users expressed interest in chronological filtering and tracking temporal changes on the clustering map for objective information consumption. Additionally, identifying intra-clusters within the conservative and liberal communities can offer a broader perspective, promoting exposure to diverse opinions while aligning with users' own beliefs.

Moreover, we acknowledge the limitations of AI models for sentiment analysis in political science, which align with general challenges in the application of AI in sociology and political science. These challenges encompass subjectivity and context dependence, emotional complexity, difficulties in recognizing sarcasm and irony, and struggles in understanding contextual nuances in text~\cite{tang2009survey, choudhury2010user, rambocas2018online, alessia2015approaches, ahire2022emotion, cui2023survey}. Specifically, sentiment analysis's subjective and context-dependent nature can lead to generalized predictions, while emotional complexity poses difficulties in capturing nuanced human emotions. Recognizing sarcasm and irony remains challenging for sentiment detection models, and understanding contextual nuances can be particularly complex for AI models.

The consideration of the importance of users' political stances and interests as well as the exploration of more appropriate labeling methods for the neutral category, are valuable directions for future research. Leveraging participants' data from social networking sites or analyzing online news consumption patterns can provide insights into users' political orientations and help refine the labeling process. We also intend to investigate how user responses to diversity-supportive interfaces vary based on the level of extremeness. Understanding how individuals with different levels of political extremism engage with and respond to interfaces that promote diverse perspectives (e.g., \system) can shed light on the effectiveness and potential challenges associated with such interfaces. These avenues present potential opportunities for future research.

\section{Conclusion}

Our study proposed a new methodology to enhance the applicability of web data for the development of an AI-based computer system. By prioritizing users' needs and characteristics, this approach holds the potential for researchers to create AI-driven interfaces that effectively address real-world social issues, such as political polarization and echo chambers. The results of our empirical user study demonstrated that \system provided participants with the flexibility to tailor their consumption of news and opinions according to their own criteria. By promoting a more balanced information consumption, our findings suggest that our approach can serve as an initial step in preventing individuals from becoming politically polarized. Our research procedures and insights can serve as a valuable resource for researchers, designers, and practitioners in the field of computer-supported collaborative work for the design of AI-based web systems to address social problems. 

\begin{acks}
This research was supported by the National Research Foundation (NRF) \& funded by the Korea government (MSIT) (No.2018R1A5A7059549, No.2021S1A5A2A03065899) and Institute of Information \& communications Technology Planning \& Evaluation (IITP) grant funded by the Korea government (MSIT) (No.2020-0-01373, Artificial Intelligence Graduate School Program (Hanyang University)).
\end{acks}

\bibliographystyle{ACM-Reference-Format}
\bibliography{Reference.bib}

\appendix
\section{Appendix}\label{appendix}

In this appendix, we provide a detailed description of the user study for modeling (Appendix~\ref{sec:appendix-userstudy}), development of our model (Appendix~\ref{sec:appendix-model}), the process of constructing the data (Appendix~\ref{sec:appendix-data}), and the additional experimental results on the reliability of political stance prediction experiments (Appendix~\ref{sec:appendix-additional-exp}).

\subsection{User Study}\label{sec:appendix-userstudy}

In this section, we present details of our user study which aims to investigate the key factors considered by real-life users when evaluating the political stance expressed in news articles. We first conducted a user study in order to investigate what factors real-world users consider and how important the factors are in determining the political stances of news articles. We provided six articles with different political stances and carefully-chosen factors to 136 respondents (69 male, 67 female) via Amazon Mechanical Turk. We asked them to respond with how important each factor is in their decision with a scale [1: not at all, 5: very much]. The "context" of a new article is the most important factor in deciding its political stance, followed by keywords, person, tone, and frequently used words. Based on this result, we observe that it is crucial (Observation 1) to learn the relationships among words/sentences to capture the context and tone of a news article, which is implicitly reflected in the article, and (Observation 2) to understand the interpretation and sentiment to real-world entities (e.g., keyword and person), which explicitly appear in a news article. 

\vspace{1mm}
\noindent

\textbf{User study setup.}

A total of 136 participants were recruited for our user study through Amazon Mechanical Turk~\footnote{\url{https://www.mturk.com/}}. Our selection process took into account diverse criteria including gender (male/female), age (ranging from between 20 to above 50), education level (ranging from high school or lower to university graduation and beyond), ethnicity (including but not limited to Caucasian, African American, Asian, Hispanic/Latino, and others), and political stance (ranging from very liberal to somewhat liberal, neutral, somewhat conservative, and very conservative). 

To investigate the influence of political stances, we carefully selected six news articles that represented different political positions. Additionally, we considered thirteen political-related factors that have been previously examined in ~\cite{lin2006side,gottopati2013learning,gentzkow2010drives,kuttschreuter2011framing,irvine1992rethinking}, These factors encompassed various aspects such as context and tone, keywords, names of individuals, topics and issues addressed, article titles, images accompanying the news articles, usage of slang words, and social and religious influences. Moreover, we focused on three news topics--- Health, Environment, and Tax--- as they are highly interconnected with political stances.

\vspace{1mm}
\noindent
\textbf{User study protocol.}

Subsequently, we proceeded with the following steps:
(1) We provided the participants with six news articles, including the title, body, and image, along with the thirteen selected factors.
(2) We asked the participants to indicate the level of significance attributed to each factor in their decision-making process, using a Likert-scale ranging from 1 (not at all significant) to 5 (very significant).
(3) We evaluated the extent to which each factor helped identify the political stance by calculating the average scores assigned by the participants.

\vspace{1mm}
\noindent
\textbf{Result and analysis.}

The study results revealed that the most significant factor in determining the political stance is the "context" of the article, followed by keywords, individuals mentioned, tone, and frequently used words. These results emphasize the importance of two key aspects: (1) understanding the interrelationships between words and sentences to capture the context and tone that are implicitly conveyed in the news article, and (2) identifying the interpretation and sentiment associated with real-world entities, such as keywords and individuals, which are explicitly present in the article.

\subsection{Model development}\label{sec:appendix-model}
We developed an AI model to predict the political stance of news articles by reflecting the results of our user study on building the model. The model consists of two key components: (1) hierarchical attention networks to learn the relationships among words/sentences in a news article with a three-level hierarchy (i.e., word-level, sentence-level, and title-level) and (2) knowledge encoding to incorporate both common and political knowledge for real-world entities, necessary for deeply understanding a news article, into the process of predicting the political stance of a news article. Regarding political knowledge, the interpretation and sentiment of even the same entities can also be different depending on their political stance. To address the subtle but important difference, we construct two knowledge graphs (KG) with different political stances, KG-lib and KG-con, and design knowledge encoding to learn to fuse the information extracted from the two political knowledge graphs. We trained the model for 50 epochs using a large dataset of news articles with five political stance labels (left, lean left, center, lean right, and right).

\subsection{Data Construction}\label{sec:appendix-data}

In this section, we describe the process to construct the datasets used in our study, namely: (1) the large-scale political news dataset, \textsf{AllSides-L}, and (2) two political knowledge graphs, \textsf{KG-lib} and \textsf{KG-con}.

\vspace{1mm}
\noindent
\textbf{\textsf{AllSides-L}.}

We complied a comprehensive political news dataset, \textsf{AllSides-L}. This dataset consists of 719,256 articles categorized into five classes: left, lean left, center, lean right, and right. The articles were collected from Allsides.com~\footnote{\url{https://www.allsides.com/}}, an American website dedicated to mitigating the adverse effects of media bias and misinformation. Allsides.com offers a wide range of political news articles that cover diverse political stances. The political classification of each news article is determined based on its associated news outlet, such as CNN or Fox. Allsides.com employs a meticulous three-step process to assign political labels: (1) domain experts label each news outlet, (2) user studies involving individuals with varying political perspectives, and (3) majority voting by individuals not involved in the user studies. Due to this rigorous labeling process, the political stances assigned by Allsides.com are generally considered as the ground truth in previous studies~\cite{feng2021kgap,zhang2022kcd}. To ensure the reliability of \textsf{AllSides-L}, we only included news articles from outlets that received high scores (7 out of 8 or higher) in the majority voting process.

\vspace{1mm}
\noindent
\textbf{\textsf{KG-lib} and \textsf{KG-con}.}

We constructed two distinct political knowledge graphs (KGs), \textsf{KG-lib} and \textsf{KG-con}, following a three-step process: (1) data collection, (2) entity/relation extraction, and (3) data cleansing. Initially, we collected a total of 496,071 posts, with 219,915 posts originating from the U.S. liberal community and 276,156 posts from the U.S. conservative community. Since these raw posts may contain numerous entities unrelated to politics, our objective was to extract political entities and their corresponding relationships from the posts. To achieve this, we employed a state-of-the-art Named Entity Recognition (NER) method~\cite{vychegzhanin2019comparison} to identify 18 political-related entities and their associated relations. Each data point was then represented as a triplet: <head entity, relation, tail entity>. To ensure the reliability of the political knowledge graphs, we manually removed any extraneous noise from the extracted triplets. As a result, we constructed two political knowledge graphs: \textsf{KG-lib} consisting of 5,581 entities and 29,967 relations, and \textsf{KG-con} comprising 6,316 entities and 33,207 relations.

\begin{table}[t!]
\centering
\caption{Comparison of the model accuracy on three real-world datasets (The bold font indicates the best results).}
\vspace{-3mm}
\label{table:model-accuracy}
\setlength\tabcolsep{8pt}
\begin{tabular}{c||c|c|c}
\toprule
\multirow{2}{*}{Method} & \multicolumn{3}{c}{Dataset} \\ 
\cmidrule(lr){2-4} 
 & \textbf{SemEval} & \textbf{AllSides-S}  & \textbf{AllSides-L} \\ 

\midrule

\textbf{Word2Vec}~\cite{mikolov2013efficient}    & 0.7027 & 0.4858 & 0.4851  \\
\textbf{GloVe}~\cite{pennington2014glove}       & 0.8071 & 0.7101 & 0.6354  \\
\textbf{ELMo}~\cite{peters-etal-2018-deep}        & 0.8678 & 0.8197 & 0.7483  \\
\textbf{BERT}~\cite{kenton2019bert}        & 0.8692 & 0.8246 & 0.7812  \\
\textbf{RoBERTa}~\cite{liu2019roberta}     & 0.8708 & 0.8535 & 0.8222  \\

\midrule

\textbf{KGAP}~\cite{feng2021kgap}    & 0.8956 & 0.8602 & N/A \\
\textbf{KCD}~\cite{zhang2022kcd}     & 0.9087 & 0.8738 & N/A \\

\midrule

\textbf{\textsf{Ours}}-RotatE & 0.9426 & 0.9151 & 0.8584 \\
\textbf{\textsf{Ours}}-HAKE & 0.9395 & 0.9216 & 0.8563 \\
\textbf{\textsf{Ours}}-ModE & \textbf{0.9521} & \textbf{0.9256} & \textbf{0.8617} \\

\bottomrule
\end{tabular}
\end{table}

\subsection{Accuracy of our AI model}\label{sec:appendix-additional-exp}
In this section, we aim to evaluate our political stance prediction model used in our empirical evaluation, with a specific focus on the accuracy of our model.

\vspace{1mm}
\noindent
\textbf{Evaluation protocol.}

To assess the accuracy of our AI model, we adopted the following evaluation protocol. We compared our model with seven baseline methods using three datasets: \textsf{SemEval}, \textsf{AllSides-S}, and \textsf{AllSides-L}: 
five text-based methods~\cite{mikolov2013efficient,pennington2014glove,peters-etal-2018-deep,kenton2019bert,liu2019roberta} and two knowledge-based political stance prediction methods~\cite{feng2021kgap,zhang2022kcd}.
Word2Vec~\cite{mikolov2013efficient}, GloVe~\cite{pennington2014glove}, and ELMo~\cite{peters-etal-2018-deep} are general language models that aim to capture context from text.
We also use pre-trained BERT~\cite{kenton2019bert} and RoBERTa~\cite{liu2019roberta} by fine-tuning on training datasets.
KGAP~\cite{feng2021kgap} is a knowledge-aware approach that leverages a political knowledge graph with graph neural networks.
KCD~\cite{zhang2022kcd}, the state-of-the-art model, considers knowledge walks from a political knowledge graph and textual cues in political stance prediction. 
To ensure consistency, 
we (1) implemented the five baseline language models using their available source code, incorporating a softmax layer for the final political stance prediction. 
Subsequently, we (2) applied the baseline methods to the \textsf{SemEval}, \textsf{AllSides-S}, and \textsf{AllSides-L} datasets, 
(3) calculated their political stance prediction accuracies, 
and (4) reported the obtained results. 
For the two knowledge-based methods (i.e., KGAP~\cite{feng2021kgap} and KCD~\cite{zhang2022kcd}), however, 
we were unable to obtain their results due to the unavailability of certain parts of the source code. 
Specifically, the source code for generating cue embeddings in KCD and the source code for generating knowledge embeddings in KGAP were not provided, respectively.
Thus, we used the results of KGAP and KCD on the \textsf{SemEval} and \textsf{AllSides-S} datasets, reported in~\cite{zhang2022kcd}. \footnote{KGAP: \url{https://github.com/BunsenFeng/news_stance_detection}, KCD: \url{https://github.com/Wenqian-Zhang/KCD}}

\vspace{1mm}
\noindent
\textbf{Results and analysis.}

We train our AI model on the three real-world political news article datasets for 50 epochs with varying knowledge embedding methods (RotatE~\cite{sun2018rotate}, HAKE~\cite{zhang2020learning}, and ModE~\cite{zhang2020learning}).
For the \textsf{SemEval} and \textsf{AllSides-L} datasets,
we apply \textit{k}-fold cross validations (10-fold for \textsf{SemEval} and 3-fold for \textsf{AllSides-S}) and then report the averaged accuracy of \textit{k}-fold cross validations.
For \textsf{AllSides-L},
we train the model on the training set (647k articles) and measure the model accuracy using the validation set (71.9k articles), which is not used in the training process.
Tables~\ref{table:model-accuracy} presents the results.
Our AI model consistently outperforms all baseline methods in terms of the model accuracy, regardless of knowledge embedding methods.
Specifically, Our model improves the state-of-the-art method, KCD~\cite{zhang2022kcd} by 4.77\% and 5.92\% in \textsf{SemEval} and \textsf{AllSides-S} datasets, respectively.
These improvements over KCD are significant,
given that KCD has already achieved quite high accuracies in those datasets. 
We also evaluate our model on a large-scale dataset (\textsf{AllSides-L}) which is $48\times$ larger and has more classes (i.e., more difficult to predict) than \textsf{AllSides-S}.
our model still significantly outperforms all baseline methods although the accuracy of our model decreases to some extent in \textsf{AllSides-L}, compared with the other datasets.
In addition, we have conducted the \textit{t}-tests with a 95\% confidence level and verified that the improvement of our model over all baseline methods are statistically significant (i.e., the \textit{p}-values are below 0.05).
As a result, considering the consistent superiority of our model to overall baseline methods, 
our experimental results convincingly demonstrate the superiority of our model in comparison to existing political stance prediction methods.

\end{document}